\documentclass[twocolumn, aps, nofootinbib, prd, floats, floatfix, amsmath, amssymb, superscriptaddress, preprintnumbers]{revtex4-2} %

\usepackage{graphicx}  
\usepackage{dcolumn}   
\usepackage{chemarrow}
\usepackage{color}
\usepackage[dvipsnames]{xcolor}
\usepackage{mathrsfs}
\usepackage{float}
\usepackage{physics}
\usepackage{slashed}
\usepackage{latexsym}
\usepackage{comment}

\usepackage[colorlinks=true, linkcolor=blue, citecolor=purple]{hyperref}
\usepackage{ulem}
\usepackage{orcidlink}
\usepackage{appendix}

\usepackage{tabularx} 
\newcolumntype{C}[1]{>{\centering\let\newline\\\arraybackslash\hspace{0pt}}m{#1}}

\begin{document}

\title{
Ultimate Quantum Precision Limit at Colliders:
Conditions and Case Studies
}

\author{Tengyu Ai}
\email{tengyuai@pku.edu.cn}
 \affiliation{School of Physics and State Key Laboratory of Nuclear Physics and Technology, Peking University, Beijing 100871, China}

\author{Qi Bi}
\email{biqii@buaa.edu.cn}
\affiliation{School of Physics, Beihang University, Beijing 100083, China}

\author{Yuxin He}
\email{yuxinhe@pku.edu.cn}
\affiliation{School of Physics and State Key Laboratory of Nuclear Physics and Technology, Peking University, Beijing 100871, China}
	
\author{Jia Liu \orcidlink{0000-0001-7386-0253}}
\email{jialiu@pku.edu.cn}
\affiliation{School of Physics and State Key Laboratory of Nuclear Physics and Technology, Peking University, Beijing 100871, China}
\affiliation{Center for High Energy Physics, Peking University, Beijing 100871, China}

\author{Xiao-Ping Wang \orcidlink{0000-0002-2258-7741}}
\email{hcwangxiaoping@buaa.edu.cn}
\affiliation{School of Physics, Beihang University, Beijing 100083, China}

\preprint{$\begin{gathered}\includegraphics[width=0.05\textwidth]{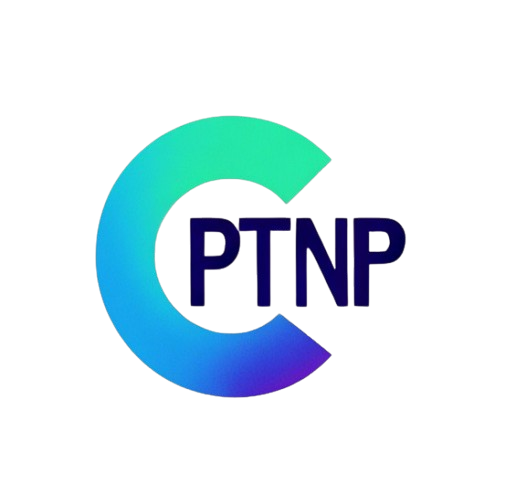}\end{gathered}$\, CPTNP-2025-019}

\begin{abstract}
We investigate whether collider experiments can reach the quantum limit of precision, defined by the quantum Fisher information (QFI), using only classical observables such as particle momenta. 
As a case study, we focus on the $\tau^+\tau^-$ system and the decay channel $\tau \to \pi \nu$, which offers maximal spin-analyzing power and renders the decay a projective measurement. 
We develop a general framework to determine when collider measurements can, in principle, saturate the QFI in an entangled biparticle system, and this framework extends naturally to other such systems. 
Within this framework, QFI saturation occurs if and only if the symmetric logarithmic derivative (SLD) commutes with a complete set of orthonormal separable projectors associated with collider-accessible measurements.
This separability condition, reflecting the independence of decay amplitudes, is highly nontrivial. To meet this condition, a key requirement is that the spin density matrix be rank-deficient, allowing the SLD sufficient freedom.
We show that the classical Fisher information asymptotically saturates the QFI for magnetic dipole moments and CP-violating Higgs interactions in selected phase-space regions, but not for electric dipole moments. These results bridge quantum metrology and collider physics, providing a systematic method to identify quantum-optimal sensitivity in collider experiments.
\end{abstract}

\maketitle

\textit{Introduction}--
The quantum Fisher information (QFI) defines the ultimate precision limit for estimating parameters encoded in quantum states, as formalized by the quantum Cramér–Rao bound \cite{Braunstein:1994zz, Paris:2008zgg, Liu:2019xfr, Safranek:2018qpw, 2020AVSQS...2a4701S}. 
For a density matrix $\rho(d)$ depending on a single parameter $d$, the QFI sets a bound on the precision of any unbiased estimator of $d$: $(\Delta d)^2 \geq 1/[N F_q(d)]$, where $N$ is the number of repeated measurements.
The QFI is intrinsically defined at the per-event level, determined solely by the density matrix of a single collider event, and thus quantifies the maximal information content carried by each copy of the state before any measurement is performed.
Once a specific measurement is performed, sensitivity is governed by the classical Fisher information (CFI), satisfying $F_c(d) \leq F_q(d)$. The QFI can be saturated if the measurement corresponds to projectors onto the eigenbasis of the symmetric logarithmic derivative (SLD), which typically involves nontrivial quantum observables~\cite{2014JPhA...47P4006T, Pang:2014xjp}.

In collider experiments, direct access to quantum states is limited; measurements are restricted to classical quantities such as particle momenta. However, spin correlations of unstable particles are preserved in the kinematics of their decay products, allowing access to the spin density matrix~\cite{Tsai:1971vv, Bernreuther:1993df, Afik:2020onf}. Techniques such as optimal observables~\cite{Atwood:1991ka, Diehl:1993br}, matrix element methods~\cite{Kondo:1988yd, D0:2004rvt, Artoisenet:2013puc} and machine learning~\cite{Cranmer:2014lly, Bahl:2021dnc} have been used to approach the CFI limit in collider measurements. Yet it remains unknown whether the quantum limit—QFI—can be reached in such classical measurement settings.

In this work, we develop a general framework to determine when quantum-optimal sensitivity can be achieved in collider experiments using classical observables. 
As a case study, we apply this framework to the entangled $\tau^+\tau^-$ system~\cite{Han:2025ewp, Fabbrichesi:2022ovb, Ehataht:2023zzt, Altakach:2022ywa, Fabbrichesi:2024wcd, Ma:2023yvd, LoChiatto:2024dmx}, with the decay $\tau \to \pi \nu$ providing maximal spin-analyzing power and yielding a rank-one projective measurement.

We show that QFI saturation occurs if and only if the SLD commutes with a complete set of separable (non-entangled) projectors for collider-accessible measurements. This separability condition, rooted in the independence of decay amplitudes, is highly nontrivial. A key requirement is that the spin density matrix be rank-deficient, allowing the SLD sufficient flexibility to align with separable eigenstates.

In $\tau^+\tau^-$ system, QFI saturation is achievable for magnetic dipole moments (MDMs) and CP-violating Higgs couplings, but not for electric dipole moments (EDMs). This framework applies to other entangled biparticle systems, such as top-quark, baryon, and gauge boson pairs.

\textit{Spin density matrix for $\tau$ pair}--
We use $\tau^+ \tau^-$ system as an example to derive the spin density matrix and demonstrate how to calculate the QFI in collider searches. 

The first scenario is $\tau^+ \tau^-$ pair from Higgs decay, $h \to \tau^+ \tau^-$ via a CP-violating phase $\delta_h$ \cite{ATLAS:2022akr, Mb:2022rxu},
\begin{align}
    \mathcal{L}_{\text{CPV}} = -\frac{m_\tau}{v} \, h \, \bar{\tau} \left[ \cos\delta_h + i \gamma_5 \sin\delta_h \right] \tau \,.
\end{align}

The other two scenarios involve $e^+e^- \to \tau^+ \tau^-$
mediated by $\tau$ MDM or EDM, which are recently measured by Refs.~\cite{Belle:2002nla, DELPHI:2003nah, OPAL:1998dsa, ARGUS:2000riz, L3:1998gov, Belle:2021ybo, ATLAS:2022ryk, CMS:2022arf, CMS:2024qjo}. 
The corresponding effective interactions with real $a_\tau$ and $d_\tau$ are
\begin{align}
    \mathcal{L}_{\text{MDM/EDM}} =\frac{1}{2} F_{\mu \nu} \, \bar{\tau} \sigma^{\mu \nu} \left[ - \frac{e}{2m_\tau}a_\tau - i \gamma_5 d_\tau\right] \tau \,.
\end{align}

For each case, we consider the spin density matrix for $\tau^+ \tau^-$ production. To avoid the integrated (or fictitious) quantum state~\cite{Afik:2020onf, Afik:2022dgh, Cheng:2023qmz}, we fix the momentum direction of the $\tau^+$, denoted by $\hat{k}$, in the center-of-mass frame of the $\tau$ pair. For an unpolarized initial state, the spin density matrix can be expressed in the helicity basis~\cite{Cheng:2023qmz} as
\begin{align}
    \rho_{\alpha\alpha'; \beta\beta'}(\hat{k}) = N_{\text{prod}}^{-1}
    \sum_{\text{ini}} M(\tau^+_\alpha \tau^-_\beta) \, M^*(\tau^+_{\alpha'} \tau^-_{\beta'}) \,,
\end{align}
where $\alpha, \beta$ label spin states, $M$ is the amplitude and $N_{\text{prod}}$ ensures $\mathrm{Tr}[\rho] = 1$. We aim to extract the small parameter $d = \delta_h, a_\tau, d_\tau$ from $\tau$ decays. Thus, $\rho$ can be expanded perturbatively with small $d$, 
\begin{align}
    \rho = \rho_0 + d \cdot \rho_1 + \mathcal{O}(d^2) \,, \label{eq:rho-expansion}
\end{align}
where $\rho_0$ is the leading contribution, and $\rho_1$ encodes the interference contribution.

\textit{Quantum Metrology and QFI}--
In quantum metrology, the ultimate precision in estimating a parameter $d$ is set by the QFI. For a state $\rho(d)$, the quantum-optimal measurement is defined by the SLD, $\hat{Q}^{\rm opt}$, which satisfies~\cite{Paris:2008zgg}
\begin{align}
    \rho_1 =  \frac{1}{2}\left\{ \rho_0, \hat{Q}^{\rm opt} \right\} \,,
    \label{eq:main:sld}
\end{align}
with the perturbative $\rho$ in Eq.~\eqref{eq:rho-expansion} and $\{\cdot, \cdot\} $ represents the anti-commutator. The QFI is:
\begin{align}
   F_q(d) =  {\rm Tr} \left[ \rho \left(\hat{Q}^{\rm opt} \right)^2 \right] \simeq {\rm Tr} \left[ \rho_0 \left(\hat{Q}^{\rm opt} \right)^2 \right]\,.
\end{align}
The quantum Cramér–Rao bound sets a lower limit on the variance of any unbiased estimator based on an observable $\hat{Q}$,
\begin{align}
    {\rm Var}\left(\hat{Q} \right) F_q(d) \geq \left( \partial_d \langle \hat{Q} \rangle_\tau \right)^2  \,,
\end{align}
where the mean value $\langle \hat{Q} \rangle_\tau$ and ${\rm Var}(\hat{Q})$ are evaluated in the spin Hilbert space of the $\tau^+\tau^-$ pair.

\textit{Generalized Quantum Measurement}--
Due to its short lifetime ($c\tau \approx 87\,\mu\mathrm{m}$), $\tau$ spin cannot be measured directly and must be inferred from its decay products. 
We focus on the two-body decay $\tau \to \pi \nu $, which offers maximal spin-analyzing power~\cite{Bernreuther:1993nd}, and the final states can be fully reconstructed.
At $e^+e^-$ colliders, the neutrino momentum typically has a two-fold ambiguity~\cite{Belle:2021ybo}. However, the precise vertex and tracking systems~\cite{Belle-II:2018jsg,K_hn_1993} enable full neutrino reconstruction using the pion’s impact parameter at Belle-II~\cite{Ehataht:2023zzt}. Machine learning techniques at the LHC also enables full $\tau$ decay kinematics ~\cite{Zhang:2025mmm}.

For single $\tau$, its decay defines a spin-space operator via its amplitude, which acts as a generalized measurement~\cite{Khan:2020seu, Qian:2020ini, Ashby-Pickering:2022umy, Wu:2024mtj}. For $\tau^- \to \pi^- \nu_\tau$, the operator is
\begin{equation}
    \hat{D}^-_{\alpha'\alpha}(\hat{q}_-) = N_D^{-1} M(\tau^-_\alpha \to \pi^-(\hat{q}_-) \nu) \, M^*(\tau^-_{\alpha'} \to \cdot\cdot ) \,,
\end{equation}
where $N_D$ is a normalization factor, and $\hat{q}_-$ is the $\pi^-$ direction in the $\tau^-$ rest frame. The phase space is denoted $\Omega_-$, with angular coordinates $\theta_-$ and $\phi_-$. The operator $\hat{D}^-(\hat{q}_-)$ is positive semidefinite, satisfies the completeness relation $\int d\Omega_- \hat{D}^-(\hat{q}_-) = 2\pi \mathbb{I}_2$, making it a legitimate measurement operator~\cite{Wu:2024mtj}. Due to maximal spin-analyzing power, $\hat{D}^-$ reduces to a rank-one projector $\ket{\hat{q}_-}\bra{\hat{q}_-}$, satisfying $(\hat{D}^-)^2 = \hat{D}^-$. A similar operator $\hat{D}^+(\hat{q}_+)$ exists for $\tau^+$. 

For $\tau^+\tau^-$ system, the decay defines a rank-one operator in the spin Hilbert space,
\begin{align}
    \hat{E}_{\alpha\alpha';\,\beta\beta'}(\hat{q}_{+},\hat{q}_-) =  \hat{D}^+_{\alpha\alpha'}(\hat{q}_+) \, \hat{D}^-_{\beta\beta'}(\hat{q}_-)  \,.
    \label{eq:decay-amplitude-operator}
\end{align}
Each $\hat{E}$ is a direct product of $\hat{D}^\pm(\hat{q}_\pm)$, associated with the pion directions $(\hat{q}_+, \hat{q}_-)$. It satisfies the properties $\hat{E}^2 = \hat{E}$ and the completeness relation $\int d\Omega_{+} d\Omega_{-} \hat{E} = 4\pi^2 \mathbb{I}_4$, see Appendix (App.) \ref{sec:app-sigma-to-E} and \ref{sec:DM-to-GMO}.

The normalized differential distribution for the process $e^{+} e^{-} \rightarrow \tau^{+} \tau^{-} \rightarrow \pi^{+} \pi^{-} \nu_\tau \bar{\nu}_\tau$, derived in App.~\ref{sec:app-sigma-to-E}, is
\begin{equation}
    \frac{d f}{d\Omega_{+} d\Omega_{-}} \equiv \frac{1}{\sigma} \frac{d\sigma}{d\Omega_{+}d\Omega_{-}} 
    = \mathrm{Tr} \left[ \frac{\hat{E}(\hat{q}_{+},\hat{q}_-)}{4\pi^2} \, \rho(\hat{k}) \right] .
    \label{eq:E-rho-f}
\end{equation}
Using Eq.~\eqref{eq:rho-expansion}, the distribution expands as
\begin{align}
    \frac{df }{d\Omega_{+} d\Omega_{-}} = \Sigma_0 + d \cdot \Sigma_1+\mathcal{O}(d^2) \,,
    \label{eq:distribution-splitting}
\end{align}
which captures the leading dependence on $d$. This directly links $\tau$ spin correlations to observed pion kinematics via projective measurements.

\textit{Optimal Observables for CFI}--
For classical measurements based on pion directions $\hat{q}_\pm$, an observable $O_d(\hat{q}_+, \hat{q}_-)$ can be constructed to estimate the parameter $d$. The statistical precision is bounded by the classical Cramér–Rao inequality~\cite{cramér1946mathematical},
\begin{align}
    \mathrm{Var}(O_d) F_c(d) \geq \left( \partial_d \langle O_d \rangle_\pi \right)^2  \,,
    \label{CR-bound}
\end{align}
where $\langle O_d \rangle_\pi$ and $\mathrm{Var}(O_d)$ are computed from the distribution in Eq.~\eqref{eq:E-rho-f}. The CFI for small $d$ is
\begin{equation}
    F_c(d) = \int d\Omega_{+} d\Omega_{-} \frac{\Sigma_1^2} {\Sigma_0}, \;
    O_d^{\mathrm{opt}}(\hat{q}_+, \hat{q}_-) = \Sigma_1 / \Sigma_0 \,,
    \label{eq:classical-opt}
\end{equation}
where $O_d^{\mathrm{opt}}$ is the classical optimal observable that saturates the classical bound Eq.~\eqref{CR-bound}. Many collider analyses aim to reach this classical limit by constructing such observables or using machine learning techniques~\cite{ParticleDataGroup:2024cfk}.

\textit{The relation between CFI and QFI}--
Each classical observable $O(\hat{q}_+ , \hat{q}_-)$ defined on pion directions corresponds to a unique quantum operator $\hat{Q}_O$ in the $\tau$ spin Hilbert space, via
\begin{align}
    \hat{Q}_O \equiv \int d\Omega_{+} d\Omega_{-} \, \hat{E}(\hat{q}_+, \hat{q}_-) \, O(\hat{q}_+, \hat{q}_-) \,.
    \label{eq:O-to-Q}
\end{align}
This construction ensures that the quantum and classical expectation values are identical using Eq.~\eqref{eq:E-rho-f}: 
\begin{align}
    \langle \hat{Q}_O \rangle_\tau = \mathrm{Tr}\left[ \rho(\hat{k}) \hat{Q}_O \right] = \langle O(\hat{q}_{+},\hat{q}_-) \rangle_\pi \,.
    \label{eq:equal-expt}
\end{align}
The variance of the classical observable also satisfies~\cite{Liu:2019xfr}
\begin{align}
    \mathrm{Var}(O) \geq \mathrm{Var}(\hat{Q}_O) \,,
\end{align}
reflecting the information loss in mapping quantum to classical observables.

Applying this to the classical optimal observable $O_d^{\mathrm{opt}}$, one obtains a hierarchy of sensitivities:
\begin{equation}
    F_q(d) \geq \frac{ \left( \partial_d \langle \hat{Q}_{O_d^{\mathrm{opt}}} \rangle_\tau \right)^2 }{ \mathrm{Var}(\hat{Q}_{O_d^{\mathrm{opt}}}) }
    \geq \frac{ \left( \partial_d \langle O_d^{\mathrm{opt}} \rangle_\pi \right)^2 }{ \mathrm{Var}(O_d^{\mathrm{opt}}) }
    = F_c(d) \,.
\end{equation}
This illustrates the general principle $F_q \geq F_c$: classical optimal observables are always less precise than the quantum-optimal operators.

\textit{Conditions to reach QFI}--
In quantum metrology, QFI saturation is achieved by performing projective measurements in the eigenbasis of the SLD,
\begin{align}
    \hat{Q}^{\rm opt} = \sum_i \lambda_i \, \hat{\Pi}_i \,,
\end{align}
where $\hat{\Pi}_i = | \psi_i \rangle \langle \psi_i |$ project onto the orthonormal eigenstates of $\hat{Q}^{\rm opt}$.

However, in $\tau$ pair decays, the collider measurement is limited to the pion momenta $q_\pm$, which determine the spin-space projector
\begin{equation}
    \hat{E}(-\hat{q}_{+},\hat{q}_-) = | \hat{q}_+ \rangle \otimes | \hat{q}_- \rangle  \langle \hat{q}_+ | \otimes \langle \hat{q}_- | 
    \equiv |E \rangle \langle E| \,,
\end{equation}
where $|\hat{q}_\pm \rangle$ are the spin states based on the pion directions in each $\tau$ rest frame. These operators $\hat{E}$ are rank-one projectors onto separable states, reflecting the independent $\tau$ decay amplitudes in Eq.~\eqref{eq:decay-amplitude-operator} (see the exclusion of decay dependence on local variables~\cite{Shi:2019kjf, BESIII:2025vsr}). Collider measurements only access such separable projectors. Given the first state $|E_1 \rangle=|\hat{q}_+, \hat{q}_-\rangle \equiv\ket{\hat{q}_+}\otimes\ket{\hat{q}_-}$, the remaining orthonormal  states $|E_j\rangle\,(j=2,3,4)$ are forced to be separable
\begin{align} 
    | -\hat{q}_+, \hat{q}_- \rangle,~~
    | \hat{q}_+, -\hat{q}_- \rangle,~~
    | -\hat{q}_+, -\hat{q}_- \rangle \,,
    \label{eq:E-projector-sets}
\end{align}
where $| \hat{q}_+, \hat{q}_- \rangle$ has components \begin{align}
\{
c_+ c_- ,
c_+ s_- e^{i \phi_-} ,
s_+ c_- e^{i \phi_+} ,
s_+ s_- e^{i (\phi_+ + \phi_-)} \},
\end{align}
where $c_\pm = \cos(\theta_\pm/2)$ and $s_\pm = \sin(\theta_\pm/2)$. The minus sign indicates that $| - \hat{q}_+\rangle$ is orthogonal to $|\hat{q}_+\rangle$ in the $\tau^+$ spin space.

This structure imposes a key constraint: if any SLD eigenstate is entangled, the QFI cannot be saturated via collider measurements. Thus, the necessary and sufficient condition for QFI saturation is that the SLD shares its eigenbasis with the accessible separable projectors, which can be expressed as
\begin{align}
    [\hat{Q}^{\rm opt}, \hat{E}_j] = 0 \quad \Leftrightarrow \quad \hat{Q}^{\rm opt} | E_j \rangle = \lambda_j | E_j \rangle  \,,
    \label{eq:eq-reach-QFI}
\end{align}
for $j = 1,2,3,4$. This means that the SLD must be simultaneously diagonalizable with all four experimental projectors, which gives four linear constraints to determine whether QFI saturation is possible.

\textit{Difficulties and rank deficit as the solution}--
The conditions in Eq.~\eqref{eq:eq-reach-QFI} require all eigenstates of $\hat{Q}^{\rm opt}$, generically written as $a \ket{\uparrow \uparrow} + b \ket{\uparrow\downarrow} + c 
\ket{\downarrow\uparrow} + d \ket{\downarrow\downarrow}$,  must be separable. This condition translates to a vanishing Schmidt rank, i.e., $ad - bc = 0$, which is a stringent constraint for biparticle system produced at colliders.

In processes like $pp \to t\bar{t}$~\cite{note1}, contributions from both quark- and gluon-initiated diagrams typically produce a full-rank mixed spin density matrix~\cite{Afik:2020onf, Bernreuther:2015yna, Cheng:2024btk}, a feature further reinforced by convolution with parton distribution functions (PDFs). In such cases, the SLD is uniquely fixed, and its eigenstates are generally entangled. Full reconstruction of $t\bar{t}$ kinematics might resolve the convolution of PDFs, but in general, achieving QFI saturation remains difficult.

However, in our perturbative setup, the rank deficiency of the spin density matrix $\rho_0$ alters the situation. Specifically, the SLD equation Eq.~\eqref{eq:main:sld}, remains unchanged under the addition of any operator in the null space of $\rho_0$, see its general form in App. \ref{sec:Form-of-SLD}. 
These components do not affect QFI, leaving $\hat{Q}^{\rm opt}$ non-unique and allowing a family of solutions.

\textit{CPV Higgs decay}--
We consider the first scenario  $h \to \tau^+ \tau^-$ , which serves as a simple and illustrative, yet counter-intuitive example. The perturbative expansion of the spin density matrix gives
\begin{align}
    \rho_0 = \frac{1}{2}
    \begin{pmatrix}
        0 & 0 & 0 & 0 \\
        0 & 1 & 1 & 0 \\
        0 & 1 & 1 & 0 \\
        0 & 0 & 0 & 0
    \end{pmatrix},
    \rho_1 = 
    \begin{pmatrix}
        0 & 0 & 0 & 0 \\
        0 & 0 & -i  & 0 \\
        0 & i  & 0 & 0 \\
        0 & 0 & 0 & 0
    \end{pmatrix},
\end{align}
with ranks 1 and 2, respectively. Both $\rho_0$ and $\rho_1$ are nonzero only in the central $2 \times 2$ block spanned by the spin states $\ket{\uparrow\downarrow}$ and $\ket{\downarrow\uparrow}$. It is thus natural to seek a solution for the SLD $Q_h$ within this subspace. A straightforward choice for $Q_h$ and the QFI are given as
\begin{align}
    Q_h = 2  \rho_1  \,, \quad F_q(\delta_h) = 4 \,.
\end{align}
It is clear that two eigenstates of $Q_h$ are entangled, so it cannot be implemented through separable projective measurements. 

Remarkably, by exploiting the rank deficiency of $\rho_0$, one can construct a modified SLD operator that does satisfy the separability requirements. The full solution is
\begin{equation}
    Q_h^{\mathrm{opt}} = \frac{2}{\sin \varphi} \begin{pmatrix}
        -\cos \varphi & 0 & 0 & e^{-i \Phi} \\
        0 & -\cos \varphi & e^{-i \varphi} & 0 \\
        0 & e^{i \varphi} & -\cos \varphi & 0 \\
        e^{i \Phi} & 0 & 0 & -\cos \varphi
    \end{pmatrix},
\end{equation}
with $\varphi \equiv \phi_1 - \phi_2$, $\Phi \equiv \phi_1 + \phi_2$, and $\phi_{1,2}$ being two free phase parameters.
$Q_h^{\rm opt}$ clearly differs from $Q_h$ , because no choice of $\phi_{1,2}$ can reduce it back to $Q_h$.
The eigenstates $|E_i\rangle$ of $Q_h^{\rm opt}$ form a complete set of orthonormal projectors, corresponding to four pion configurations with fixed angular directions in the $\tau$ rest frames, summarized in Table~\ref{tab:h-cpv-pion-direction}. In each case, the $\pi^\pm$ are emitted in the transverse plane, perpendicular to $\hat{k}$.

\begin{table}[h!]
    \centering
   \begin{tabular}{| C{1.5cm} |C{1.2cm}|C{1.2cm}|C{1.2cm}|C{1.2cm}|}
        \hline
        CPV $h$  & $\theta_+$ & $\theta_-$ & $\phi_+$ & $\phi_-$ \\ \hline
        $|E_1 \rangle$   & $\pi/2$ & $\pi/2$ & $\phi_1$         & $\phi_2$   \\ \hline
        $|E_2 \rangle$   & $\pi/2$ & $\pi/2$ & $\phi_1 + \pi$   & $\phi_2$   \\ \hline
        $|E_3 \rangle$   & $\pi/2$ & $\pi/2$ & $\phi_1$         & $\phi_2+\pi$  \\ \hline
        $|E_4 \rangle$   & $\pi/2$ & $\pi/2$ & $\phi_1 + \pi$   & $\phi_2 + \pi$  \\ \hline
    \end{tabular}
    \caption{Four optimal projective measurements for CPV Higgs decay and the $\pi^\pm$ directions are perpendicular to tau directions.}
    \label{tab:h-cpv-pion-direction}
\end{table}

\textit{MDM and EDM of $\tau$}--
For the process $e^+ e^- \to \tau^+ \tau^-$ at low-energy lepton colliders with unpolarized beams, the calculations of density matrices $\rho_{0,1}$ and QFI for MDM and EDM scenarios are detailed in App. \ref{sec:Explicit-results-for-tau}.

For the MDM case, the SLD satisfying Eq.\eqref{eq:eq-reach-QFI} is diagonal in the helicity basis.
The corresponding projective measurements are the four orthonormal product states $\ket{\uparrow\uparrow}$, $\ket{\uparrow\downarrow}$, $\ket{\downarrow\uparrow}$, and $\ket{\downarrow\downarrow}$. These are experimentally accessible by selecting pion momentum configurations aligned or anti-aligned with the $\tau$ momentum $\hat{k}$ in the rest frame. Notably, these directions do not have the two-fold ambiguity for neutrinos. 
The four measurement settings are listed in Table~\ref{tab:MDM-pion-direction}.

\begin{table}[h!]
    \centering
    \begin{tabular}{|C{2.3cm}|C{1.cm}|C{1cm}|C{1.cm}|C{1.cm}|}
        \hline
        MDM  & $\theta_+$ & $\theta_-$ & $\phi_+$ & $\phi_-$ \\ \hline
        $|E_1\rangle = \ket{\downarrow\downarrow}$ & $\pi$ & $\pi$ & $-$ & $-$ \\ \hline
        $|E_2\rangle = \ket{\downarrow\uparrow}$ & $\pi$ & $0$   & $-$ & $-$ \\ \hline
        $|E_3\rangle = \ket{\uparrow\downarrow}$ & $0$   & $\pi$ & $-$ & $-$ \\ \hline
        $|E_4\rangle = \ket{\uparrow\uparrow}$ & $0$   & $0$   & $-$ & $-$ \\ \hline
    \end{tabular}
    \caption{Four projective measurements for MDM, with pion directions along the $\hat{k}$ axis. 
    }
    \label{tab:MDM-pion-direction}
\end{table}

In contrast, for the EDM scenario, we find no SLD that satisfies Eq.~\eqref{eq:eq-reach-QFI}. To systematically explore possible SLDs in both MDM and EDM cases, we proposed and utilized the amplitude-based method, which is equivalent to the matrix-based method, see App. \ref{sec:Form-of-SLD}.

\begin{figure}[htb]
  \centering
  \includegraphics[width=0.95\linewidth]{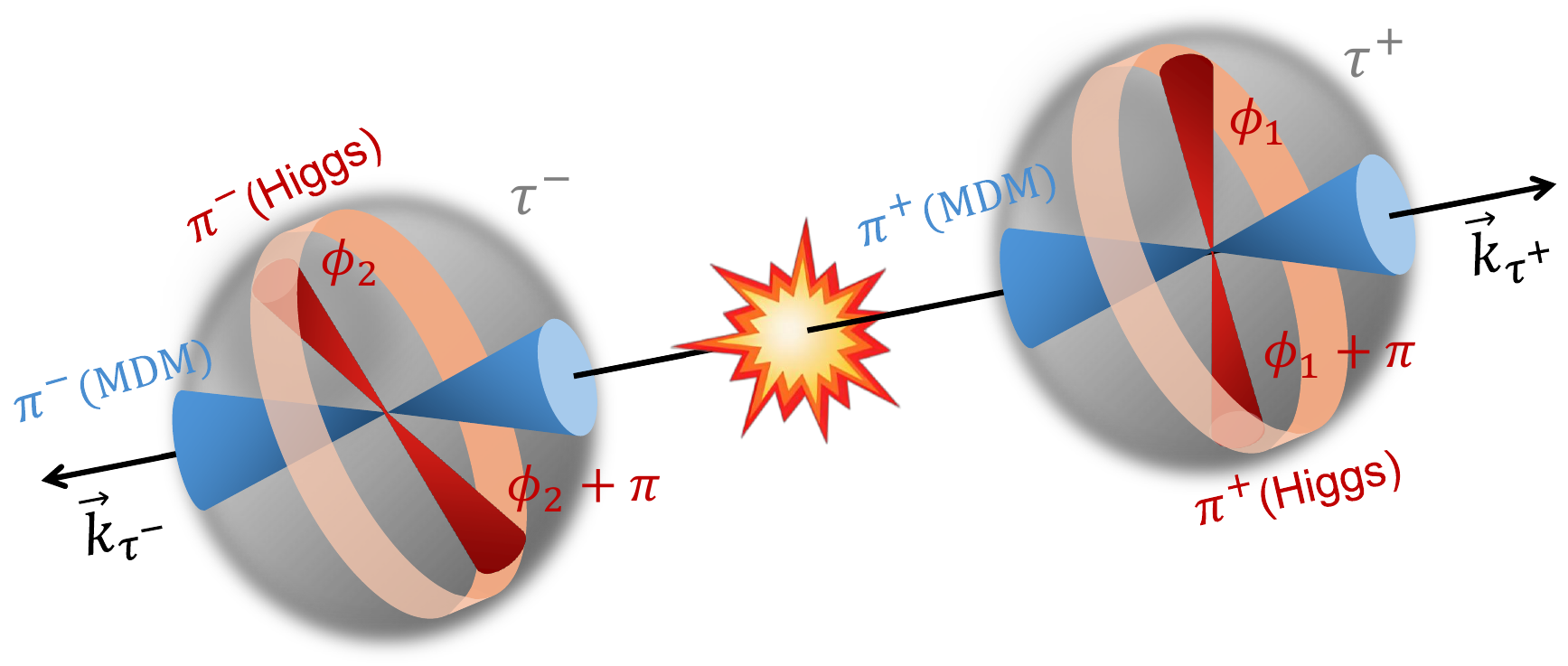}
  \caption{Illustration of the optimal conical phase space for the MDM and Higgs cases. For the MDM case, the phase space includes four fixed $\pi^\pm$ direction configurations (Tab.~\ref{tab:MDM-pion-direction}), each expanded into a cone of opening angle $\delta\theta_\pi$, denoted as Cone. For the Higgs case, the optimal directions vary with the free azimuthal parameters $\phi_{1,2}$, effectively forming a belt around the equator widened by $\delta\theta_\pi$, denoted as Belt.}
  \label{fig:optimal-phase-space}
\end{figure}

\textit{Approaching QFI in collider measurements}--
While projective measurements using the operators $\hat{E}_{1,2,3,4}$ can, in principle, saturate the QFI for the MDM and Higgs cases, the corresponding pion directions occupy zero-measure phase space, yielding no observable events. To overcome this, we define quasi-QFI measurements by expanding each optimal direction into a cone of angular size $\delta\theta_\pi$, recovering the ideal limit as $\delta\theta_\pi \to 0$. We compute the CFI within these cones using Eq.\eqref{eq:classical-opt}. The optimal conical phase space is illustrated in Fig.~\ref{fig:optimal-phase-space}.

\begin{figure*}[htb]
  \centering
  \includegraphics[width=0.40\linewidth]{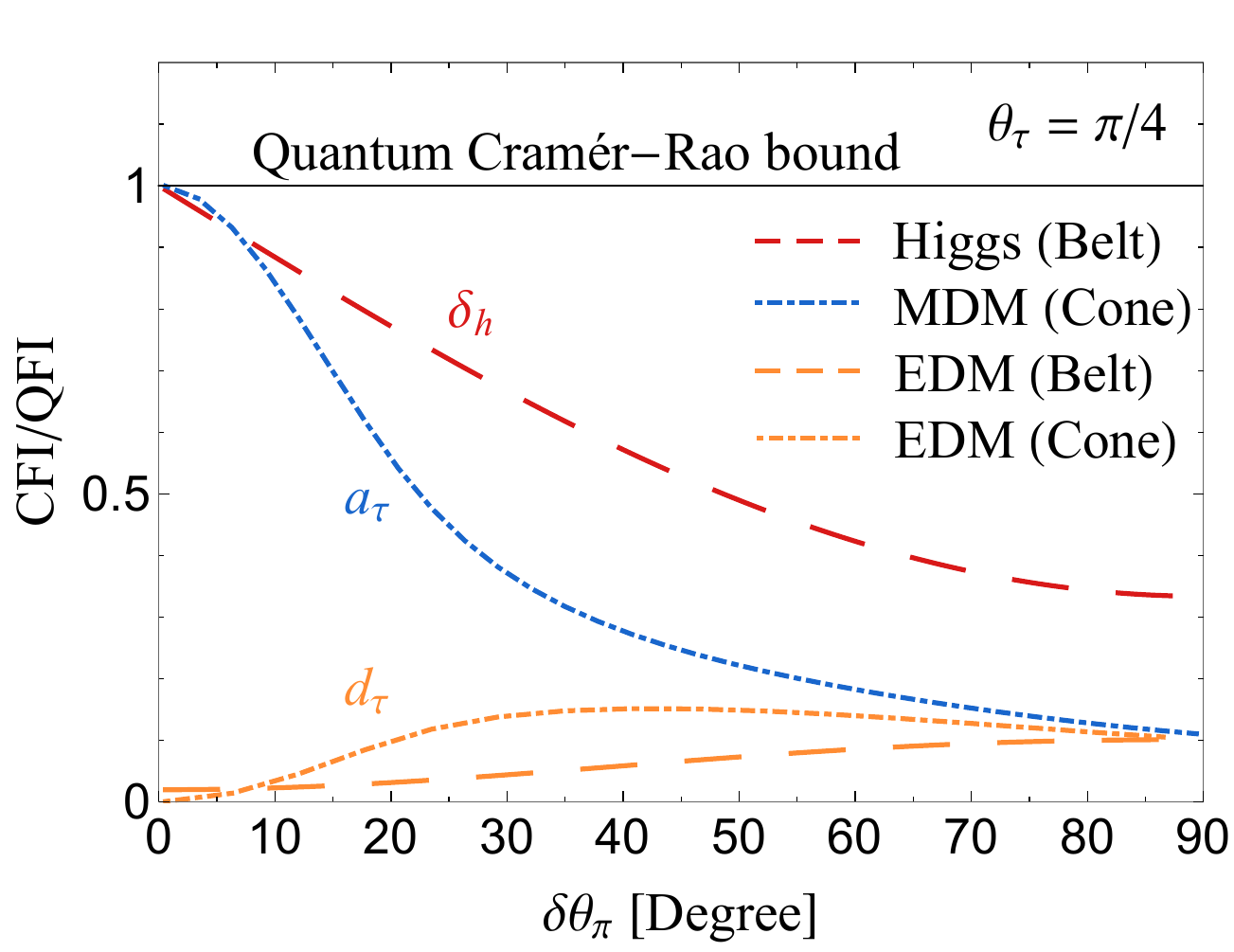}
 ~~~~~~~~
\includegraphics[width=0.40\linewidth]{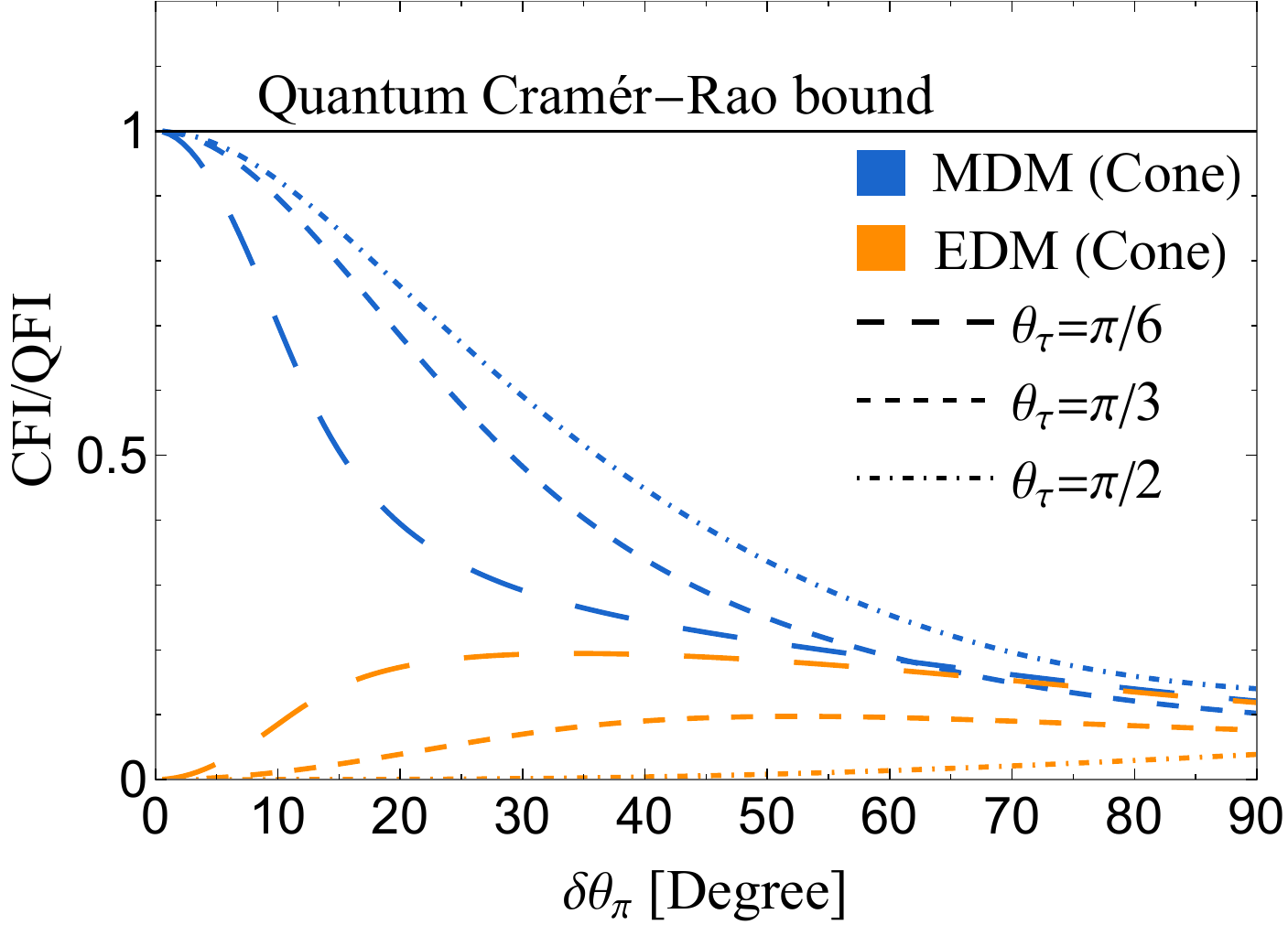}
  \caption{
  CFI as a function of cone size $\delta \theta_\pi$ in restricted phase space, normalized to the corresponding QFI for MDM and Higgs cases. The EDM case is shown for comparison using the same restricted phase space. The $\tau^+$ direction is fixed at polar angle $\theta_{\tau} = \pi/4$ for MDM and EDM cases in the left panel. In the right panel, $\theta_\tau$ is fixed as $\pi/6$, $\pi/3$, and $\pi/2$, respectively. 
  }
  \label{fig:QFIvsCFI}
\end{figure*}

Fig.~\ref{fig:QFIvsCFI} shows the calculated CFI for the MDM with different choice of $\theta_\tau$, the polar angle of $\tau$, while the Higgs case is $\theta_{\tau}$ independent.
Both converge to the QFI as $\delta \theta_\pi \to 0$. This is analytically confirmed in App. \ref{sec:CFI-to-QFI}, which proves that the phase-space–restricted CFI asymptotically reaches the QFI. These results highlight that conventional CFI methods can achieve the quantum precision limit in collider settings. For the EDM case, no separable projective measurements can saturate the QFI, but its CFI and QFI are compared in the same way for completeness.

Except $\tau \to \pi \nu$ decay, other decay channels have reduced spin-analyzing power and correspond to non-projective positive operator-valued measures (POVM). In these cases, the resulting CFI is strictly lower, and saturation of the QFI is not possible within our current framework, as shown in App. \ref{sec:non-maximal-spin-analyzing-power}. We restrict to the projective case in this work and defer general POVM effects to future studies.

\textit{Discussions and Conclusions}--
We developed a general framework to determine when collider experiments, limited to classical observables, can achieve the quantum precision limit set by the QFI. Using the $\tau^+ \tau^-$ system as a benchmark, we showed that collider-accessible measurements correspond to projective operators onto separable spin states, due to the independent two decay amplitudes. QFI saturation requires that the SLD commute with all such separable projectors—a nontrivial condition made possible by the rank deficiency of the spin density matrix, which grants the SLD sufficient flexibility.

Applying this framework to three benchmark processes, we found that QFI saturation is achievable for the MDM and CP-violating Higgs cases, but only within restricted regions of phase space. For the EDM case, no such separable SLD exists. We also demonstrate that the conventional CFI method, when confined to suitable phase-space regions, can asymptotically approach the QFI. CP violation does not preclude QFI saturation, as seen in the Higgs example. While our main discussion considers $\tau$ pairs with any fixed $\hat{k}$, statistics from different $\hat{k}$ can be combined by averaging $F_q(\hat{k})$ with weight $\sigma^{-1} d\sigma/d\hat{k}$, which leaves our conclusions unchanged.

While our analysis focuses on the $\tau^+\tau^-$ system, the framework naturally extends to a broad class of biparticle systems in which spin information is reconstructed from decay kinematics.
Examples include  top-quark pairs (e.g., produced by $e^+e^-$ collisions), baryon pairs (e.g., $\Delta$ baryons, hyperons, and heavy-flavor baryons), and gauge boson pairs (e.g., $WW$, $ZZ$), either produced directly or via heavy meson decays.
Extensions to generalized POVMs--relevant for decays with reduced spin-analyzing power--will be explored in future work.

It is worth emphasizing that QFI saturation is achievable only in specific regions of collider phase space; full-phase-space measurements cannot saturate the QFI everywhere but can nevertheless yield greater overall precision because of their larger statistical power. 

In this sense, the per-event QFI is a property of the underlying entangled production state and provides a measurement-independent benchmark above the CFI obtainable from any specific differential cross section. While this benchmark does not itself improve the sensitivity of an already-recorded data set with a fixed measurement strategy, it allows one to diagnose how close a given analysis is to the per-event quantum limit, to assess whether further methodological refinements are warranted given the required experimental resources, and to identify decay modes and regions of phase space that are intrinsically most informative about the parameter of interest, thereby guiding future studies that incorporate realistic detector effects, backgrounds, and analysis choices.

\vspace{12pt}
\textit{Acknowledgements}.
We would like to thank Zhicheng Yang for helpful discussions.
The work of J.L. is supported by the National Science Foundation of China under Grant No. 12235001, No. 12475103 and State Key Laboratory of Nuclear Physics and Technology under Grant No. NPT2025ZX11.
The work of X.P.W. is supported by the National Science Foundation of China under Grant No. 12375095, and the Fundamental Research Funds for the Central Universities. J.L. and X.P.W. also thank APCTP, Pohang, Korea, for their hospitality during the focus program [APCTP-2025-F01], from which this work greatly benefited. J.L. and X.P.W. also thank the Mainz Institute for Theoretical Physics (MITP) of the PRISMA+ Cluster of Excellence (Project ID 390831469) for its  hospitality and support during the completion of this work. The authors gratefully acknowledge the valuable discussions and insights provided by the members of the Collaboration of Precision Testing and New Physics.

\onecolumngrid

\clearpage

\appendix

\begin{center}
{\large \bf Appendix \\
Ultimate Quantum Precision Limit at Colliders: Conditions and Case Studies\\
}

\end{center}

This appendix provides detailed derivations, extended calculations, and mathematical proofs supporting the results presented in the main text.

\tableofcontents

\section{From Differential Cross Section to Generalized Measurement Operators}
\label{sec:app-sigma-to-E}

In collider experiments, quantum states composed of fermion pairs are routinely produced, offering a natural setting for quantum measurement. As in other quantum systems, extracting information from these states requires performing a series of measurements. Collider processes, in particular, offer built-in measurement opportunities through particle decays, which serve as generalized quantum measurements. As demonstrated in previous studies~\cite{Bernreuther:1993nd}, the complete spin density matrix of the fermion pair is encoded in the kinematic distributions of their decay products.

A representative and experimentally relevant case is the $\tau^+ \tau^-$ system, which forms a two-qubit quantum state of spin-1/2 particles. $\tau$ pairs can be produced either directly or via the decay of heavier particles, such as the Higgs boson. Although direct spin measurements of $\tau$ leptons are not feasible at colliders, their spin information is imprinted in the momenta of their decay products through channels like $\tau^\pm \to \pi^\pm \overset{\scriptscriptstyle(-)}{\nu}$.

At an $e^+e^-$ collider, we consider the process where the initial electron-positron pair undergoes production and decay interactions: $e^+ e^- \rightarrow \tau^+(k_+) + \tau^-(k_-) \rightarrow \pi^+(q_+) + \pi^-(q_-) + X$, where $X$ denotes the undetected neutrinos. The corresponding cross section can be written as:
\begin{align}
    \sigma = \frac{1}{4E_e^2|\Delta v|} \int \frac{d^3q_+}{(2\pi)^3 2E_+} \frac{d^3q_-}{(2\pi)^3 2E_-}  \int_X d\text{PS}_X (2\pi)^4 \delta^{(4)}(p_{\text{ini}} - q_+ - q_- - q_X) 
     |\overline{\mathcal{M}}(\text{ini} \rightarrow \tau^+ \tau^- \rightarrow \pi^+ \pi^- + X)|^2,
    \label{cross:origin}
\end{align}
where $|\overline{\mathcal{M}}|^2$ is the amplitude square averaged over the initial-state spins. Since the $\tau^+ \tau^-$ state is unstable, the matrix element can be factorized into production and decay parts:
\begin{align}
    \mathcal{M}(\text{ini} \rightarrow \tau^+ \tau^- \rightarrow \pi^+ \pi^- + X) = & \sum_{\alpha, \beta} \mathcal{M}(\text{ini} \rightarrow \tau^+_\alpha(k_+) \tau^-_\beta(k_-)) \times \frac{i}{k_+^2 - m_{\tau}^2 + i m_{\tau} \Gamma_{\tau}} \frac{i}{k_-^2 - m_{\tau}^2 + i m_{\tau} \Gamma_{\tau}} \nonumber \\
    &\times \mathcal{M}(\tau^+_\alpha(k_+) \rightarrow \pi^+ \bar{\nu}) \mathcal{M}(\tau^-_\beta(k_-) \rightarrow \pi^- \nu),
\end{align}
where $\alpha,\beta$ represent spin indices for $\tau^+$ and $\tau^-$, and $\Gamma_{\tau}$ is the total decay width of $\tau$. Using the narrow-width approximation, the $\tau$ propagators can be factorized, simplifying the full process into a production followed by decay.

Substituting this into Eq.~\eqref{cross:origin} and applying the identity $\lim_{\epsilon \to 0} \frac{\epsilon}{\epsilon^2 + x^2} = \pi \delta(x)$, one can integrate over $X$ and re-express the remaining phase space using: (i) $\tau^\pm$ momenta in the center-of-mass frame, and (ii) $\pi^\pm$ momenta in the respective $\tau^\pm$ rest frames: 
\begin{align}
    \sigma & =   \frac{1}{4E_e^2|\Delta v|} \int\frac{d^3q_+}{(2\pi)^3 2E_+} \frac{d^3q_-}{(2\pi)^3 2E_-}  \int_X d\text{PS}_X (2\pi)^4 \delta^{(4)}(p_{\text{ini}} - q_+ - q_- - q_X)  \left( \frac{\pi}{m \Gamma_\tau} \right)^2 \delta(k_+^2 - m_{\tau}^2) \delta(k_-^2 - m_{\tau}^2) \nonumber \\
    & \times \left|\sum_{\alpha,\beta} \mathcal{M}(e^+ e^- \rightarrow \tau^+_\alpha(k_+) \tau^-_\beta(k_-)) \times \mathcal{M}(\tau^+_\alpha(k_+) \rightarrow \pi^+ \bar{\nu}) \mathcal{M}(\tau^-_\beta(k_-) \rightarrow \pi^- \nu)\right|^2 \nonumber \\
    = & N_\sigma \int d\Omega_{\tau}\, d\Omega_+ d\Omega_-  |\overline{\mathcal{M}}(\hat{k})|^2  \sum_{\alpha, \alpha'; \beta, \beta'} \rho_{\alpha\alpha'; \beta\beta'}(\hat{k}) \hat{D}^+_{\alpha'\alpha}(\hat{q}_+) \hat{D^-}_{\beta'\beta}(\hat{q}_-),
    \label{corsssection:origin}
\end{align}
where $d\Omega_{\pm}$ are the solid angles of the $\pi^\pm$ directions $\hat{q}_{\pm}$ in their respective $\tau^\pm$ rest frames, $d\Omega_{\tau}$ is the solid angle of the $\tau^+$ direction $\hat{k}$ in the center-of-mass frame, $\Gamma_\tau$ is the total decay width of $\tau$ and $X$ represents $\nu$ and $\bar\nu$.
$|\overline{\mathcal{M}}(\hat{k})|^2$ is the amplitude square averaged over the initial-state spins for the tau pair production process without decay.
$N_\sigma$ is a prefactor from kinematics and phase space integration, which ensures the cross-section $\sigma$ to be the product of total production cross-section $\sigma(e^+e^- \to \tau^+ \tau^-)$ times decay branching ratio squared $[\text{BR}(\tau \to \pi \nu)]^2$.

The spin density matrix $\rho(\hat{k})$ captures all spin information of the $\tau$ pair at a given $\hat{k}$. The spin density matrix $\rho$, and the decay matrix $\hat{D}$ for single $\tau$ are given by:

\begin{align}
    \rho_{\alpha\alpha'; \beta \beta'}(\hat{k}) =& \frac{1}{|\overline{\mathcal{M}}|^2} \overline{\sum_{\text{initial}}}   \mathcal{M}(\text{initial} \rightarrow \tau^+_\alpha(\hat{k}) \tau^-_\beta(-\hat{k})) \mathcal{M}^*(\text{initial} \rightarrow \tau^+_{\alpha'}(\hat{k}) \tau^-_{\beta'}(-\hat{k})),  \\
    \hat{D}^\pm_{\alpha\alpha'}(\hat{q}_{\pm}) = & \frac{f_\pi^2}{2m_\tau^3|q|}\mathcal{M}_\pm(\hat{q}_\pm,\alpha') \mathcal{M}_\pm(\hat{q}_\pm,\alpha),
    \label{eq:single measurement2}
\end{align}
where $\overline{\sum}_{\text{initial}}$ denotes averaging over initial spin states, $\mathcal{M}_+(\hat{q}_+,\alpha) = \mathcal{M}\!\left(\tau^{+}_\alpha(k_+) \to \pi^{+}(\hat{q}_+)\bar{\nu}\right)$ and similarly for $\mathcal{M}_-$. After integrating over the rest of the decay phase space, the dependence of $\hat{D}(\hat{q}_{\pm})$ reduces to the direction $\hat{q}_\pm$ alone. It can be explicitly written as:
\begin{equation}
    \hat{D}^\pm_{\alpha\alpha'}(\hat{q}_{\pm}) =\frac{1}{2}(\mathbb{I}_2\mp \hat{q}_{\pm}\cdot \vec{\sigma})_{\alpha\alpha'},
    \label{eq:single-measurement}
\end{equation}
where $\mathbb{I}_2$ is the $2 \times 2$ identity matrix, $\vec{\sigma}$ denotes the Pauli matrices, and the “$\mp$" sign reflects the opposite spin-analyzing powers of the decays $\tau^+ \to \pi^+ \bar{\nu}$ and $\tau^- \to \pi^- \nu$, for which $\hat{D}$ exhibits maximal spin-analyzing power.

The operator $\hat{D}$ related with $\tau$ decay satisfies the following properties: 
\begin{align}
\text{Tr}[\hat{D}^\pm] = 1, \quad (\hat{D}^\pm)^2 = \hat{D}^\pm, \quad \int d\Omega_{\pm}  \hat{D}^\pm(\hat{q}_\pm) = 2\pi \mathbb{I}_2,
\label{eq:D-properties}
\end{align}
confirming that it is a legitimate quantum measurement operator.

By combining the decay matrices, we define a Hermitian measurement operator $\hat{E}(\hat{q}_{+},\hat{q}_{-})$ acting on the $\tau$ pair Hilbert space:
\begin{equation}
    \hat{E}_{\alpha\alpha';\beta\beta'}(\hat{q}_{+},\hat{q}_{-})=\hat{D}^+_{\alpha\alpha'}(\hat{q}_{+})\otimes\hat{D}^-_{\beta\beta'}(\hat{q}_{-})=\frac{1}{4}(\mathbb{I}_{4}-\hat{q}_{+}\cdot \vec{\sigma}\otimes\mathbb{I}_2+ \mathbb{I}_2\otimes\vec{\sigma}\cdot\hat{q}_{-}-\hat{q}_{+}\cdot \vec{\sigma}\otimes\vec{\sigma}\cdot\hat{q}_{-})_{\alpha\alpha';\beta\beta'},
    \label{measure}
\end{equation}
where $(A\otimes B)_{\alpha\alpha'; \beta\beta'} = A_{\alpha\alpha'} B_{\beta\beta'}$. Using this operator, the total cross section in Eq.~\eqref{corsssection:origin} can be written as:
\begin{align}
    \sigma = &  N_\sigma \int d\Omega_{\tau}\, d\Omega_+ d\Omega_-  |\overline{\mathcal{M}}(\hat{k})|^2  \sum_{\alpha, \alpha'; \beta, \beta'} \hat{E}_{\alpha'\alpha;\beta'\beta}(\hat{q}_{+},\hat{q}_{-}) \rho_{\alpha\alpha'; \beta\beta'}(\hat{k})
    \nonumber \\
    = & N_\sigma \int d\Omega_{\tau}\, d\Omega_+ d\Omega_-  |\overline{\mathcal{M}}(\hat{k})|^2 \cdot \mathrm{Tr}[\hat{E}(\hat{q}_{+},\hat{q}_{-}) \rho(\hat{k})].
\end{align}
The normalized double-differential angular distribution of the pions can be defined as
\begin{equation}
    \frac{df}{d\Omega_{+}d\Omega_{-}}( \hat{k};\hat{q}_{+},\hat{q}_{-})=\frac{1}{4\pi^2}\mathrm{Tr}[\hat{E}(\hat{q}_{+},\hat{q}_{-}) \rho(\hat{k})] \,,
    \label{dis:tau}
\end{equation}
which describes the angular correlations of the pion directions $\hat{q}_\pm$ in the respective $\tau$ rest frames, given that the $\tau^+$ travels along direction $\hat{k}$ in the lab frame. This normalized distribution encodes the spin-correlation information of the $\tau^+ \tau^-$ pair.

Thus, the decay process $\tau^{\pm} \rightarrow \pi^{\pm} \bar{\nu}(\nu)$ can be interpreted as a generalized quantum measurement on the spin density matrix $\rho(\hat{k})$ of the $\tau^+\tau^-$ system, implemented by a set of positive operator-valued measures (POVMs) $\hat{E}(\hat{q}_{+}, \hat{q}_{-})$. These operators form a complete set of projectors~\cite{Wu:2024mtj}, satisfying:
\begin{align}
    \text{Tr}[\hat{E}] = 1, \quad
    \hat{E}^2(\hat{q}_{+},\hat{q}_{-})=\hat{E}(\hat{q}_{+},\hat{q}_{-}), \quad \frac{1}{4\pi^2}\int d\Omega_{+} d\Omega_{-} \hat{E}(\hat{q}_{+},\hat{q}_{-})= \mathbb{I}_{4}.
\end{align}
It is straightforward to verify that each $\hat{E}(\hat{q}_{+}, \hat{q}_{-})$ is a rank-one projector on the joint spin Hilbert space of the $\tau$ pair:
\begin{align}
    \hat{E}(-\hat{q}_{+},\hat{q}_{-})=\ket{\hat{q}_+,\hat{q}_-}\bra{\hat{q}_+,\hat{q}_-},
    \label{eq:measure-E-diparticle}
\end{align}
where $\ket{\hat{q}_{+}, \hat{q}_{-}} = \ket{\hat{q}_{+}} \otimes \ket{\hat{q}_{-}}$, and each spin-1/2 state $\ket{\hat{q}_{\pm}} = \{\cos(\theta_\pm/2), \sin(\theta_\pm/2) e^{i\phi_\pm}\}$ corresponds to spin polarization along the direction $\hat{q}_{\pm}$ in the respective $\tau$ rest frames.

Consequently, the measurements of collider experiments are restricted to accessing separable spin states of the form $\ket{\hat{q}_{+}, \hat{q}_{-}}$.

\section{From Density Matrix to Generalized Measurement Operators }
\label{sec:DM-to-GMO}

Consider the production of a $\tau^+\tau^-$ pair in a collider, with momenta $\pm \hat{k}$ for $\tau^+$ and $\tau^-$, respectively, in their center-of-mass frame.
The spin density matrix of the produced pair can be written as
\begin{equation}
    \rho^{\rm prod}_{\tau^+ \tau^-}  = \sum_{\alpha,\beta;\alpha',\beta'} 
    c_{\alpha\alpha';\beta\beta'} 
    \,|\tau_{\alpha}^+,\tau_{\beta}^- \rangle
    \langle \tau_{\alpha'}^+,\tau_{\beta'}^-| \,,
\end{equation}
where $|\tau_{\alpha}^+,\tau_{\beta}^- \rangle$ denotes the $\tau^+\tau^-$ biparticle state, with $\alpha$ and $\beta$ labeling the helicities. The momenta are omitted for brevity.

The subsequent $\tau^+\tau^-$ decays are described by the $S$-matrix, which produces various possible final states.
Selecting a specific decay channel $a$ corresponds to a quantum channel that maps the $\tau^+\tau^-$ system into the Hilbert space of that channel:
\begin{align}
     \rho^{\rm final}_{\pi^+ \pi^-}
     &= \frac{K_{a}\,\rho^{\rm prod}_{\tau^+\tau^-}\,K_{a}^\dagger}
             {{\rm Tr}\!\left[\rho^{\rm prod}_{\tau^+\tau^-}K_{a}^\dagger K_{a}\right]}
     = \frac{1}{{\rm norm}}\,K_{a}\,\rho^{\rm prod}_{\tau^+\tau^-}\,K_{a}^\dagger \,,
     \label{eq:quantum-channel-from-decay}
\end{align}
For concreteness, let us specify the decay channel $a$ to be $\pi\nu$. In this case $K_a$ is the Kraus operator associated with channel $a=\pi\nu$, whose matrix elements coincide with those of the $S$-matrix:
\begin{align}
    \langle \pi^+\bar{\nu},\pi^-\nu| K_{\pi\nu} |\tau_{\alpha}^+,\tau_{\beta}^- \rangle
    &= \langle\pi^+\bar{\nu},\pi^-\nu| S(\tau_{\alpha}^+\tau_{\beta}^- \to \pi^+\bar{\nu}\pi^-\nu) |\tau_{\alpha}^+,\tau_{\beta}^- \rangle
     \nonumber\\
    &\propto \mathcal{M}\!\left(\tau^{+}(\hat{k},\alpha) \to \pi^+\bar{\nu}\right)
       \mathcal{M}\!\left(\tau^{-}(-\hat{k},\beta) \to \pi^-\nu\right) \,,
\end{align}
and the Kraus operator $K_{\pi\nu}$ can be written explicitly as
\begin{align}
    K_{\pi\nu}  = \int d\Omega_+\, d\Omega_-  \mathcal{M}\!\left(\tau^{+}(\hat{k},\alpha) \to \pi^+(\hat{q}_+)\bar{\nu}\right)
       \mathcal{M}\!\left(\tau^{-}(-\hat{k},\beta) \to \pi^-(\hat{q}_-)\nu\right) 
       | \hat{q}_+ , \hat{q}_- \rangle \langle \tau_{\alpha}^+,\tau_{\beta}^-|
\end{align}
where $\hat{q}_{\pm}$ are the pion momenta. 

Thus, for the decay $\tau \to \pi \nu$, the helicity density matrix $\rho^{\rm prod}_{\tau^+ \tau^-}$ is mapped into a new density matrix in the Hilbert space of pion momentum. Explicitly,
\begin{align}
    \rho^{\rm final}_{\pi^+ \pi^-} 
    &= \frac{1}{{\rm norm}} 
   \sum_{\alpha,\beta,\alpha',\beta'} 
   \int d\Omega_+\, d\Omega_- \;
   \mathcal{M}_+(\hat{q}_+,\alpha) \,
   \mathcal{M}_-(\hat{q}_-,\beta) \, \left(\rho^{\rm prod}_{\tau^+ \tau^-} \right)_{\alpha\beta ;\alpha'\beta'} \nonumber \\
   &\quad\times 
   \int d\Omega_+^{\prime}\, d\Omega_-^{\prime} \;
   \mathcal{M}^*_+(\hat{q}'_+,\alpha') \,
   \mathcal{M}^*_-(\hat{q}'_-,\beta')\, | \hat{q}_+ , \hat{q}_- \rangle \langle \hat{q}_+^{\prime} , \hat{q}_-^{\prime}| \,,
   \label{eq:new-density-matrix}
\end{align}
where we have introduced the shorthand $\mathcal{M}_+(\hat{q}_+,\alpha) = \mathcal{M}\left(\tau^{+}(\hat{k},\alpha) \to \pi^{+}(\hat{q}_+)\right)$, and similarly for $\mathcal{M}_-$.
Equation~\eqref{eq:new-density-matrix} describes the state of the system after decay into the specific final state ($\pi\nu$). This process can be understood as a quantum channel. 

The decay products are then measured in the collider, where the four-momenta of the final particles in the $\tau \to \pi\nu$ channel are detected. Such a measurement can be represented by a projective operator
\begin{equation}
    \hat{L}_{\pi\nu}(\hat{q}_\pi) = |\hat{q}_\pi\rangle \langle \hat{q}_\pi| \,,
\end{equation}
so that the probability of obtaining a final state with momenta $\{\hat{q}_+,\hat{q}_-\}$ is
\begin{align}
    \mathcal{P}(\hat{q}_+,\hat{q}_-)& = \frac{1}{\sigma}\frac{d\sigma}{ d\Omega_+ d \Omega_-}(\hat{q}_+,\hat{q}_-) = \langle \hat{q}_{\pi^+} \hat{q}_{\pi^-} | \rho^{\rm final}_{\pi^+\pi^-} | \hat{q}_{\pi^+} \hat{q}_{\pi^-}\rangle ={\rm Tr}\left[ \rho^{\rm final}_{\pi^+\pi^-} \left(\hat{L}_{\pi^+}(\hat{q}_+)\otimes\hat{L}_{\pi^-}(\hat{q}_-)\right)\right] 
    \label{eq:probability-form}\\
    &=\frac{1}{\rm norm}{\rm Tr}\left[\rho^{\rm prod}_{\tau^+\tau^-} K_{\pi\nu}^\dagger \left(\hat{L}_{\pi^+}(\hat{q}_+)\otimes\hat{L}_{\pi^-}(\hat{q}_-)\right)K_{\pi\nu}\right] \,,
\end{align}
where, in the last line, Eq.~\eqref{eq:quantum-channel-from-decay} has been used. 
Next, we show that the Kraus operator and the momentum projector combine to
\begin{align}
    K_{\pi\nu}^\dagger \left(\hat{L}_{\pi^+}(\hat{q}_+)\otimes\hat{L}_{\pi^-}(\hat{q}_-)\right)K_{\pi\nu} 
    &= \mathcal{M}^*_+(\hat{q}_+,\alpha') \mathcal{M}^*_-(\hat{q}_-,\beta') 
   \mathcal{M}_+(\hat{q}_+,\alpha) \mathcal{M}_-(\hat{q}_-,\beta) 
      | \tau_{\alpha'}^+,\tau_{\beta'}^- \rangle \langle \tau_{\alpha}^+,\tau_{\beta}^{-}|\nonumber\\
      &=\left(\frac{2m_\tau^3|q|}{f_\pi^2}\right)^2\hat{D}^+_{\alpha'\alpha}(\hat{q}_{+})\hat{D}^-_{\beta'\beta}(\hat{q}_{-})| \tau_{\alpha'}^+,\tau_{\beta'}^- \rangle \langle \tau_{\alpha}^+,\tau_{\beta}^{-}| \,,
\end{align}
which serves as an operator acting on the spin Hilbert space. 
The normalization factor ``norm" is given by
\begin{align}
    {\rm norm} & = {\rm Tr}\!\left[\rho^{\rm prod}_{\tau^+\tau^-}K_{\pi\nu}^\dagger K_{\pi\nu}\right] \nonumber \\
    &= \sum_{\alpha,\beta;\alpha',\beta'}\int d\Omega_+\, d\Omega_- \;
       \mathcal{M}_+(\hat{q}_+,\alpha)
       \mathcal{M}^*_+(\hat{q}_+,\alpha')
       \left(\rho^{\rm prod}_{\tau^+ \tau^-} \right)_{\alpha\beta ;\alpha'\beta'}
       \mathcal{M}_-(\hat{q}_-,\beta)
       \mathcal{M}_-^*(\hat{q}_-,\beta')\nonumber\\
       &=\left(\frac{2m_\tau^3|q|}{f_\pi^2}\right)^2\sum_{\alpha,\beta;\alpha',\beta'}\int d\Omega_+d\Omega_-\hat{D}^+_{\alpha'\alpha}(\hat{q}_{+})
       \left(\rho^{\rm prod}_{\tau^+ \tau^-} \right)_{\alpha\beta ;\alpha'\beta'} \hat{D}^-_{\beta'\beta}(\hat{q}_{-})=4\pi^2\left(\frac{2m_\tau^3|q|}{f_\pi^2}\right)^2 \,,
\end{align}
where in the last line we used Eq.~\eqref{eq:single measurement2} and Eq.~\eqref{eq:D-properties}. It should also be noted that the normalization factor ``norm" is independent of $\rho$, ensuring that the quantum channel operation remains linear in the density matrix.

Thus, the probability in Eq.~\eqref{eq:probability-form} can be rewritten with a positive operator-valued measurement $\hat{E}$, which defines a generalized quantum measurement on $\tau^+\tau^-$ helicity Hilbert space, 
\begin{align}
    \mathcal{P}(\hat{q}_+,\hat{q}_-) = \frac{1}{\sigma}\frac{d\sigma}{ d\Omega_+ d \Omega_-}(\hat{q}_+,\hat{q}_-) = \langle \hat{q}_{\pi^+} \hat{q}_{\pi^-} | \rho^{\rm final}_{\pi^+\pi^-} | \hat{q}_{\pi^+} \hat{q}_{\pi^-}\rangle
    = {\rm Tr}\left[\rho^{\rm prod}_{\tau^+\tau^-} \frac{\hat{E}(\hat{q}_+,\hat{q}_-)}{4\pi}\right]
    \label{eq:probability-for-state-and-E}
\end{align}
 with
\begin{equation}
    \hat{E}(\hat{q}_+,\hat{q}_-)=\hat{D}^+(\hat{q}_{+})\otimes\hat{D}^-(\hat{q}_{-}) = \frac{4 \pi}{\rm norm} K_{\pi\nu}^\dagger \left(\hat{L}_{\pi^+}(\hat{q}_+)\otimes\hat{L}_{\pi^-}(\hat{q}_-)\right)K_{\pi\nu}\,,
    \label{eq:E-and-momentum-operator}
\end{equation}
which brings us back to Eq.~\eqref{measure} in Sec. \ref{sec:app-sigma-to-E}.

Eqs.~\eqref{eq:probability-for-state-and-E} and \eqref{eq:E-and-momentum-operator} establish a rigorous correspondence between the two measurement pictures. Specifically, the momentum–projector measurement on the daughter particles,
$\hat{L}_{\pi^+}(\hat{q}_+) \otimes \hat{L}_{\pi^-}(\hat{q}_-)$, collapses the final state with a probability identical to that obtained by measuring the mother $\tau^+\tau^-$ helicity state with the operator $\hat{E}(\hat{q}_+,\hat{q}_-)$. This demonstrates that \textit{measuring the momenta of the decay products is mathematically equivalent to performing a generalized quantum measurement on the helicity of the mother particles. }

Finally, the above construction can be directly extended to an arbitrary decay channel $a$ by introducing a new $\hat{D}^a_{\alpha\alpha'}$, which takes the same form as Eq.~\eqref{eq:single-measurement} but with the spin–analyzing power appropriate to channel $a$.
\\

\section{General and Collider-Accessible Forms of the SLD}
\label{sec:Form-of-SLD}

Once the spin density matrix and the parameter of interest are defined, the corresponding symmetric logarithmic derivative (SLD) can be computed. For a rank-deficient spin density matrix $\rho_0$, the SLD is not uniquely determined. Instead, there exists a family of operators that satisfy the SLD equation and yield the same quantum Fisher information (QFI). As shown in~\cite{Liu:2019xfr}, the general form of an operator $\hat{Q}$ that satisfies the SLD equation can be written as: 
\begin{equation}
    \hat{Q} = \sum_{\substack{i,j\\ p_i + p_j \neq 0}} \frac{\langle p_i | \rho_1 | p_j \rangle}{p_i + p_j} |p_i\rangle \langle p_j| + \sum_{\substack{i,j\\ p_i = p_j = 0}}r_{ij} |p_i\rangle \langle p_j|,
\end{equation}
where $p_i$ and $|p_i\rangle$ are the semi-positive eigenvalues and corresponding eigenstates of $\rho_0$, respectively. The coefficients $r_{ij}$ are arbitrary complex numbers subject to the Hermiticity condition $r_{ij} = r_{ji}^*$. The second term corresponds to the null space of $\rho_0$, which is nontrivial due to its rank deficiency and provides the flexibility in defining the SLD.

The eigenstates of $\hat{Q}$ determine the optimal measurement basis that saturates the QFI bound. However, collider experiments are restricted to measurements constructed from decay products, which correspond to separable states in the tensor-product Hilbert space of the two particles. As a result, not all SLD eigenstates are experimentally accessible. To determine whether the optimal SLD can be implemented in collider settings, we introduce two complementary approaches:
\begin{enumerate}
    \item a \textit{matrix-based method}, which is linear-algebraic and direct;
    \item an \textit{amplitude-based method}, which leverages particle physics amplitudes and provides physical intuition.
\end{enumerate}

\subsection{Matrix-Based Method}
\label{sec:Matrix-Based-method}

We focus on the spin Hilbert space in analyzing measurement separability. For a single spin-$1/2$ particle, the separability condition is trivially satisfied. The corresponding projective measurement is a spin projection along a direction $\hat{\alpha}$, described by the operator:
\begin{equation}
    \hat{\Pi}^{\text{single}}=\frac{1}{2}\left(\mathbb{I}+\hat{\alpha}\cdot\vec{\sigma}\right)=\ket{\hat{\alpha}}\bra{\hat{\alpha}},
\end{equation}
where $\hat{\alpha}$ is a unit vector that defines the measurement direction in spin space (i.e., the direction along which the spin is projected), and $\vec{\sigma}$ is the Pauli matrix. To perform the optimal measurement, the states $\ket{\hat{\alpha}}$ should be chosen as the eigenstates of the SLD operator $\hat{Q}$.

For a single-particle system, the decay projector has the same form as Eq.~\eqref{eq:single-measurement}. Choosing a specific state $\ket{\hat{\alpha}}$ corresponds to choosing a specific direction $\hat{q}_\pm$ of the decay product, with $\ket{\hat{\alpha}} = \ket{\hat{q}}$, which can always be achieved. Thus, for the single-particle case, the optimal measurement saturating the QFI bound can always be realized.

For a biparticle system, the corresponding measurement takes the form given in Eq.~\eqref{eq:measure-E-diparticle}. In this case, the separability condition becomes more intricate. To ensure that the SLD is experimentally implementable in a collider setting, its eigenstates must be non-entangled. This requirement is equivalent to the commutativity condition:
\begin{align}
[\hat{Q}, \hat{E}_j] = 0 \quad \Leftrightarrow \quad \hat{Q} | E_j \rangle = \lambda_j | E_j \rangle,
\label{eq:app:eq-reach-QFI}
\end{align}
where $\hat{E}_j$ is a separable measurement operator and $|E_j\rangle$ is a corresponding separable eigenstate. Directly solving this eigenvalue problem is feasible but computationally challenging. However, the analysis can be simplified by expanding $\hat{Q}$ in the Pauli basis:
\begin{equation}
    \hat{Q} = a \mathbb{I}_4 +\sum_{i}b_i^+(\sigma_i\otimes \mathbb{I}_2)+\sum_{j}b_j^-(\mathbb{I}_2\otimes \sigma_j ) + \sum_{i,j} c_{ij} \, \sigma_i \otimes \sigma_j,
\end{equation}
with 16 real coefficients from $a$, $b_i^{\pm}$, and $c_{ij}$. If the operator $\hat{Q}$ is separable, it can be decomposed in the Pauli basis as follows, 
\begin{align}
\hat{Q} = P_1 \left(\mathbb{I}_2 \otimes \mathbb{I}_2\right) + P_2 \left(\hat{q}_+ \cdot \vec{\sigma}\right) \otimes \mathbb{I}_2 + P_3 \mathbb{I}_2 \otimes \left(\hat{q}_- \cdot \vec{\sigma}\right) + P_4 \left(\hat{q}_+ \cdot \vec{\sigma}\right) \otimes \left(\hat{q}_- \cdot \vec{\sigma}\right),
\end{align}
where the expansion coefficients are then given by
\begin{equation}
    a = P_1 \,, \qquad \vec{b}^\pm = P_{2/3} \, \hat{q}_\pm \,, \qquad c_{ij} = P_4 \, \hat{q}_+^i \hat{q}_-^j \,.
\end{equation}
This formulation implies that the matrix $c = [c_{ij}]$ must either be an outer product of two $\hat{q}_\pm$ vectors or identically zero. When $ P_{2/3} \neq 0 $, the relation $ c_{ij} \propto b^+_i b^-_j = (\vec{b}^+ \vec{b}^{-T})_{ij} $ must hold.  Separable solutions exist in the scenarios involving MDM and CP-violating Higgs couplings, but no such solution is feasible in the case of the EDM.

\subsection{Amplitude-Based Method}
\label{sec:Amp-Based-method}

The amplitude-based method systematically explores possible SLDs that can be constructed from separable measurement operators. For a scattering process, the spin density matrix $\rho_d$ can be theoretically computed from the transition amplitudes $\mathcal{M}$ as:
\begin{equation}
    \rho_{d} = \frac{\mathcal{M} \rho_{\text{int}} \mathcal{M}^\dagger}{A},
\end{equation}
where $\rho_{\text{int}}$ is the spin density matrix of the initial state, and $A = \mathrm{Tr}(\mathcal{M} \rho_{\text{int}} \mathcal{M}^\dagger)$ is a normalization factor. This result follows by expanding $\mathcal{M}$ and $A$ linearly in the perturbative parameter $d$:
\begin{align}
    \mathcal{M}(d) &= \mathcal{M}_0 + d \cdot \mathcal{M}_1 + \mathcal{O}(d^2), \nonumber \\
    A(d) &= A_0 + d \cdot A_1 + \mathcal{O}(d^2), \nonumber
\end{align}
where
\begin{equation}
A_0=\mathrm{Tr}\left[\rho_{\text{int}}\mathcal{M}^{\dagger}_0\mathcal{M}_0\right],\quad A_1=\mathrm{Tr}\left[\rho_{\text{int}}(\mathcal{M}^{\dagger}_0\mathcal{M}_1+\mathcal{M}^{\dagger}_1\mathcal{M}_0) \right].
\label{eq:A}
\end{equation}
Expanding the density matrix linearly yields $\rho_0$ and $\rho_1$, and the SLD equation becomes
\begin{equation}
    \left( \frac{1}{2} \hat{Q} \mathcal{M}_0 - \mathcal{M}_1 + \frac{A_1}{2 A_0} \mathcal{M}_0 \right) \rho_{\text{int}} \mathcal{M}_0^\dagger 
    + \mathcal{M}_0 \rho_{\text{int}} \left( \frac{1}{2} \hat{Q}^\dagger \mathcal{M}_0 - \mathcal{M}_1 + \frac{A_1}{2 A_0} \mathcal{M}_0 \right)^\dagger = 0.
    \label{eq:sld:scattering}
\end{equation}
A sufficient condition for Eq.~\eqref{eq:sld:scattering} to hold is existing a Hermitian $\hat{Q}$ that satisfies:
\begin{equation}
    \frac{1}{2} \hat{Q} \mathcal{M}_0 -\tilde{\mathcal{M}}_1=0.
    \label{eq:sufficienta}
\end{equation}
where $\tilde{\mathcal{M}}_1 \equiv (\mathcal{M}_1- \frac{A_1}{2 A_0} \mathcal{M}_0)$.
Notably, when the initial state $\rho_{\text{int}}$ is full rank, this sufficient condition can also become necessary under certain conditions, thereby establishing the equivalence:
\begin{equation}
    \text{for full-rank $\rho_{\text{int}}$,  Eq.~\eqref{eq:sld:scattering}} 
    \quad \Leftrightarrow \quad 
    \text{Eq~\eqref{eq:sufficienta}}, 
    \quad \text{with the conditions:} \quad 
    \begin{cases}
        (1)\; \mathcal{M}_1 \mathcal{M}_0^{\mathrm{MP}} \mathcal{M}_0 = \tilde{\mathcal{M}}_1 \\
        (2)\; \mathcal{M}_0^\dagger \mathcal{M}_1 = \mathcal{M}_1^\dagger \mathcal{M}_0
    \end{cases},
    \label{eq:sufficient}
\end{equation} 
where $\mathcal{M}_0^{\mathrm{MP}}$ denotes the Moore-Penrose pseudoinverse of $\mathcal{M}_0$. This equivalence is indeed satisfied in all three cases discussed in App.~\ref{sec:Explicit-results-for-tau}, and its mathematical proof is provided in App.~\ref{sec:proof-of-equivalence}.

The first condition corresponds to the solvability criterion for a linear system of the form $\frac{1}{2} X \mathcal{M}_0 = \mathcal{M}_1$. The operator $\mathcal{M}$ can be interpreted as a map from the initial spin space $\mathcal{H}^{\text{int}}$ to the final spin space $\mathcal{H}^{\text{fin}}$. The product $\mathcal{M}_0^{\mathrm{MP}} \mathcal{M}_0$ acts as a projector onto the support of $\mathcal{M}_0$, meaning the kernel of $\mathcal{M}_0$ must be a subset of the kernel of $\mathcal{\mathcal{M}}_1$. This condition is satisfied in MDM and Higgs case because, at tree level, the kernel of $\mathcal{M}$ is determined by the helicity structure of the $e$--$e$--$\gamma$ vertex for MDM. While for the Higgs case, the initial spin space collapses to one-dimension. The vertex structure is the same for both $\mathcal{M}_0$ and $\mathcal{M}_1$, implying that they share the same kernel.

The condition (2) guarantees the existence of a Hermitian solution. This requirement depends on the specific interaction and scattering process, and it is fulfilled in our cases.

Notably, the rank deficiency of the density matrix $\rho_0$ is reflected by the non-zero kernel of $\mathcal{M}_0^\dagger$. This introduces a degree of freedom, enabling the selection of an appropriate $\hat{Q}$ that satisfies the non-entangled eigenstate condition.

From the amplitude-based method, Eq.~\eqref{eq:sufficient} provides an alternative method to solve the SLD equation. Suppose $\xi$ is the eigenvector of the SLD optimal operator $\hat{Q}$ with eigenvalue $\lambda$, then it yields:
\begin{equation}
    \xi^\dagger (\lambda ~\mathcal{M}_0^i - \tilde{\mathcal{M}}_1^i) = 0, \quad \text{for} \quad i = 1, \dots, n,
\end{equation}
where $\mathcal{M}_0^i$ and $\tilde{\mathcal{M}}_1^i$ are the $i$-th column vectors of $\mathcal{M}_0$ and $\tilde{\mathcal{M}}_1$ respectively, and $n$ is the total column number corresponding to the dimension of initial spin space $\text{dim}(\mathcal{H}^{int})$. Since $\hat{Q}$ is Hermitian, its eigenvalue $\lambda$ is guaranteed to be real, which implies that
\begin{equation}
    \lambda = \frac{\xi^\dagger \tilde{\mathcal{M}}_1^1}{\xi^\dagger \mathcal{M}_0^1} = \frac{\xi^\dagger \tilde{\mathcal{M}}_1^2}{\xi^\dagger \mathcal{M}_0^2} = \dots = \frac{\xi^\dagger \tilde{\mathcal{M}}_1^n}{\xi^\dagger \mathcal{M}_0^n}.
    \label{eq:sld:eigen}
\end{equation}
To obtain a non-entangled eigenstate, $\xi$ must take the form
\begin{equation}
    \xi(\hat{q}_+, \hat{q}_-) = \begin{pmatrix} \cos \frac{\theta_+}{2} \\ \sin \frac{\theta_+}{2} e^{i \phi_+} \end{pmatrix} 
    \otimes 
    \begin{pmatrix} \cos \frac{\theta_-}{2} \\ \sin \frac{\theta_-}{2} e^{i \phi_-} \end{pmatrix}
    = \begin{pmatrix} 
\cos \frac{\theta_+}{2} \cos \frac{\theta_-}{2} \\
\cos \frac{\theta_+}{2} \sin \frac{\theta_-}{2} e^{i \phi_-} \\
\sin \frac{\theta_+}{2} \cos \frac{\theta_-}{2} e^{i \phi_+} \\
\sin\frac{\theta_+}{2} \sin \frac{\theta_-}{2} e^{i (\phi_+ + \phi_-)}
\end{pmatrix} \,,
\end{equation}
with $\theta_\pm$ and $\phi_\pm$ as the angular coordinates for the unit vector $\hat{q}_\pm $. 
Substituting this form into Eq.~\eqref{eq:sld:eigen}, 
one can determine the optimal angles for $\hat{q}_\pm$ in collider measurements.

\section{Explicit results for CPV-Higgs, MDM and EDM}
\label{sec:Explicit-results-for-tau}
In this section, we compute the spin density matrices and SLD operators for three distinct scenarios: CP-violating Higgs (CPV Higgs) interactions, magnetic dipole moments (MDM), and electric dipole moments (EDM) for the $\tau$ lepton. All calculations are performed in the helicity basis \(\{\hat{\mathbf{n}},\hat{\mathbf{r}},\hat{\mathbf{k}}\}\)  within the center-of-mass frame of the \(\tau\)-pair, where the basis vectors are defined as follows:
\begin{equation}
    \hat{\mathbf{n}}=\frac{1}{\sin\theta}(\hat{\mathbf{p}}\times\hat{\mathbf{k}}),\quad \hat{\mathbf{r}}=\frac{1}{\sin\theta}(\hat{\mathbf{p}}-\cos\theta\hat{\mathbf{k}}),
\end{equation}
where \(\hat{\mathbf{k}}\) is the direction of \(\tau^+\), \(\hat{\mathbf{p}}\) is the direction of \(e^+\) or $h$ (depending on the scenario), and \(\theta\) is the production angle of \(\tau^+\) satisfying \(\cos\theta=\hat{\mathbf{p}}\cdot\hat{\mathbf{k}}\).
The spin density matrices are calculated analytically at leading order. Higher-order corrections are known to be small, e.g. $O(\alpha)$ in $e^+e^- \to \tau^+ \tau^-$\cite{Krinner:2021cqs} and $h \to \tau^+ \tau^-$\cite{Han:2017yhy}. Since both the QFI and CFI depend on the density matrix, their corrections should be of the same small order. Therefore, higher-order effects are neglected in this initial study.

\subsection{CPV-Higgs}
We begin by outlining the steps for deriving the SLD operator and the QFI in the CPV Higgs scenario. To facilitate the QFI analysis, we first compute the spin density matrix. The spin density matrix for the two $\tau$ leptons in the CPV Higgs scenario is given by:
\begin{align}
    \rho = \frac{1}{2}
    \begin{pmatrix}
        0 & 0 & 0 & 0 \\
        0 & 1 & e^{-2i\delta_h} & 0 \\
        0 & e^{2i\delta_h} & 1 & 0 \\
        0 & 0 & 0 & 0
    \end{pmatrix}.
\end{align}
For the state $\rho(\delta_h)$, the SLD operator $\hat{Q}$ associated with the quantum-optimal measurement for estimating the CP-violating phase parameter $\delta$ can be obtained using either the matrix method or the amplitude-based method discussed above.
Solving of the SLD equation, we obtain the general form of the SLD operator:
\begin{equation}
\hat{Q} = \begin{pmatrix}
    r_{44} & -\frac{r_{43}}{\sqrt{2}} & \frac{r_{43}}{\sqrt{2}} & r_{42} \\
    -\frac{r_{34}}{\sqrt{2}} & \frac{r_{33}}{2} & -\frac{r_{33}}{2} - 2i & -\frac{r_{32}}{\sqrt{2}} \\
    \frac{r_{34}}{\sqrt{2}} & -\frac{r_{33}}{2} + 2i & \frac{r_{33}}{2} & \frac{r_{32}}{\sqrt{2}} \\
    r_{24} & -\frac{r_{23}}{\sqrt{2}} & \frac{r_{23}}{\sqrt{2}} & r_{22}
\end{pmatrix}.
\end{equation}
Here, the parameters $r_{ij}$ are complex and satisfy the Hermiticity condition $r_{ij} = r_{ji}^*$. 
In the Pauli basis, the components of $\hat{Q}$ are given by
\begin{align}
\vec{b}^{\, +} &= \left\{
    \frac{-r_{23} - r_{32} + r_{34} + r_{43}}{\sqrt{2}},
    \frac{i (r_{23} - r_{32} - r_{34} + r_{43})}{\sqrt{2}},
    r_{44} - r_{22}
\right\}, \nonumber \\
\vec{b}^{\, -} &= \left\{
    \frac{r_{23} + r_{32} - r_{34} - r_{43}}{\sqrt{2}},
    -\frac{i (r_{23} - r_{32} - r_{34} + r_{43})}{\sqrt{2}},
    r_{44} - r_{22}
\right\}, \nonumber \\
c &= \begin{pmatrix}
    r_{24} - r_{33} + r_{42} & -i r_{24} + i r_{42} - 4 & \frac{r_{23} + r_{32} + r_{34} + r_{43}}{\sqrt{2}} \\
    -i r_{24} + i r_{42} + 4 & -r_{24} - r_{33} - r_{42} & -\frac{i (r_{23} - r_{32} + r_{34} - r_{43})}{\sqrt{2}} \\
    -\frac{r_{23} + r_{32} + r_{34} + r_{43}}{\sqrt{2}} & \frac{i (r_{23} - r_{32} + r_{34} - r_{43})}{\sqrt{2}} & r_{22} - r_{33} + r_{44}
\end{pmatrix}. \nonumber
\end{align}

Applying the amplitude-based method, we calculate the SLD operator explicitly in the Higgs rest frame, where the spin space is reduced to one dimension due to the zero spin of the Higgs boson. The transition amplitude $M$ is given by
\begin{equation}
    \mathcal{M} = 2 y k \begin{pmatrix}
0 \\
1 - i \delta_h \\
1 + i \delta_h \\
0
\end{pmatrix}= 2 y k \begin{pmatrix} 0 \\ 1 \\ 1 \\ 0 \end{pmatrix} + \delta_h \cdot 2 y k \begin{pmatrix} 0 \\ -i \\ i \\ 0 \end{pmatrix}
\equiv \mathcal{M}_0+\delta_h \cdot \mathcal{M}_1,
\end{equation}
where $ k $ is the momentum magnitude of the $ \tau^\pm $, $y = m_\tau/v$ with $v$ being the vacuum expectation value of the Higgs field. We find that the normalization factor $A = 8k^2 y^2$ is independent of $\delta_h$ at leading order. As a result, the SLD equation in Eq.~\eqref{eq:sld:scattering} simplifies since $A_1 = 0$.

Since the Higgs spin space is one-dimensional, Eq.~(\ref{eq:sld:eigen}) simplifies considerably, with the sole requirement that the eigenvalue $\lambda$ be real. This leads to the condition:
\begin{equation}
    \text{Im}(\lambda) = \text{Im} \left( \frac{\xi^\dagger \mathcal{M}_1}{\xi^\dagger \mathcal{M}_0} \right) = 0,
\end{equation}
which leads to:
\begin{equation}
    2 \operatorname{Re} \left( \frac{1}{1 + \cot \left( \frac{\theta_1}{2} \right) \tan \left( \frac{\theta_2}{2} \right) e^{i (\phi_1 - \phi_2)}} \right) = 1.
\end{equation}
Consider the orthonormal basis $ \xi(\pm \hat{q}_+, \pm \hat{q}_-) $, we can obtain the results in Table.~\ref{tab:h-cpv-pion-direction}, with the eigenvalues for the SLD being:
\begin{equation}
    \lambda = 2\tan \left( \frac{\phi_1 - \phi_2}{2} \right), 
    \quad -2\cot \left( \frac{\phi_1 - \phi_2}{2} \right), 
    \quad -2\cot \left( \frac{\phi_1 - \phi_2}{2} \right), 
    \quad 2\tan \left( \frac{\phi_1 - \phi_2}{2} \right),
\end{equation}
where $\phi_{1,2}$ are the two free azimuthal angle parameters.
Finally, the separable SLD operator is simply given by
\begin{equation}
    Q_h^{\mathrm{opt}} =  \frac{2}{\sin \varphi} \begin{pmatrix}
        -\cos \varphi & 0 & 0 & e^{-i \Phi} \\
        0 & -\cos \varphi & e^{-i \varphi} & 0 \\
        0 & e^{i \varphi} & -\cos \varphi & 0 \\
        e^{i \Phi} & 0 & 0 & -\cos \varphi
    \end{pmatrix},
\end{equation}
with $ \varphi = \phi_1 - \phi_2 $ and $ \Phi = \phi_1 + \phi_2 $.

\subsection{MDM}

In the MDM scenario, we derive the SLD operator and QFI starting with the spin density matrix $\rho$, expanded to first order in the MDM parameter 
$d = a_\tau$. The unperturbed spin density matrix $\rho_0$ is determined by the electromagnetic interaction, takes the form:
\begin{equation}
    \rho_0 = c_0 \begin{pmatrix}
        \frac{1}{2} (\cos 2\theta + 3) & -\frac{i m}{\sqrt{s}} \sin 2\theta & -\frac{i m}{\sqrt{s}} \sin 2\theta & -\sin^2\theta \\
        \frac{i m}{\sqrt{s}} \sin 2\theta & \frac{4 m^2}{s} \sin^2\theta & \frac{4 m^2}{s} \sin^2\theta & \frac{i m}{\sqrt{s}} \sin 2\theta \\
        \frac{i m}{\sqrt{s}} \sin 2\theta & \frac{4 m^2}{s} \sin^2\theta & \frac{4 m^2}{s} \sin^2\theta & \frac{i m}{\sqrt{s}} \sin 2\theta \\
        -\sin^2\theta & -\frac{i m}{\sqrt{s}} \sin 2\theta & -\frac{i m}{\sqrt{s}} \sin 2\theta & \frac{1}{2} (\cos 2\theta + 3)
    \end{pmatrix},
    \label{eq:rho0:etau}
\end{equation}
where $ c_0 = \left(8 m^2 \sin^2\theta/s + \cos 2\theta + 3\right)^{-1} $ with $s$ being the square of the center-of-mass energy of the $\tau$ pair. To ease the notation, we denote the mass $m_\tau$ and the polar angle $\theta_\tau$ of the outgoing $\tau^+$ as $m$ and $\theta$, respectively.

To account for the leading-order correction in the MDM parameter $d$, we retain only linear terms and obtain the corresponding matrix $\rho_1$ as:
\begin{equation}
    \rho_1 = (1 - \frac{4 m^2}{s}) c_0^2 \begin{pmatrix}
        -2 \sin^2 \theta (\cos 2\theta + 3) & -\frac{i c_1 \sqrt{s}}{4 m} \sin 2\theta & -\frac{i c_1 \sqrt{s}}{4 m} \sin 2\theta & 4 \sin^4 \theta \\
        \frac{i c_1 \sqrt{s}}{4 m} \sin 2\theta & 2 \sin^2 \theta (\cos 2\theta + 3) & 2 \sin^2 \theta (\cos 2\theta + 3) & \frac{i c_1 \sqrt{s}}{4 m} \sin 2\theta \\ 
        \frac{i c_1 \sqrt{s}}{4 m} \sin 2\theta & 2 \sin^2 \theta (\cos 2\theta + 3) & 2 \sin^2 \theta (\cos 2\theta + 3) & \frac{i c_1 \sqrt{s}}{4 m} \sin 2\theta \\
        4 \sin^4 \theta & -\frac{i c_1 \sqrt{s}}{4 m} \sin 2\theta & -\frac{i c_1 \sqrt{s}}{4 m} \sin 2\theta & -2 \sin^2 \theta (\cos 2\theta + 3)
    \end{pmatrix},
\end{equation}
where \( c_1 = -\frac{8 m^2}{s} \sin^2 \theta + \cos 2\theta + 3 \).

To construct the SLD operator, we apply the amplitude-based method. Starting with the scattering amplitude, we expand it linearly in the MDM parameter $d$ as:
\begin{equation}
    \mathcal{M} = \begin{pmatrix}
        i (1 + d) \cos^2 \frac{\theta}{2} & 0 & 0 & i (1 + d) \sin^2 \frac{\theta}{2} \\
        -\frac{\sqrt{s}}{4 m} \left(\frac{4 m^2}{s} + d\right) \sin\theta & 0 & 0 & \frac{\sqrt{s}}{4 m} \left(\frac{4 m^2}{s} + d\right) \sin\theta \\
        -\frac{\sqrt{s}}{4 m} \left(\frac{4 m^2}{s} + d\right) \sin\theta & 0 & 0 & \frac{\sqrt{s}}{4 m} \left(\frac{4 m^2}{s} + d\right) \sin\theta \\
        -i (1 + d) \sin^2 \frac{\theta}{2} & 0 & 0 & -i (1 + d) \cos^2 \frac{\theta}{2}
    \end{pmatrix}
    =\mathcal{M}_0+d\cdot \mathcal{M}_1,
\end{equation}
where the zeroth- and first-order terms are:
\begin{equation}
    \mathcal{M}_0=\begin{pmatrix}
        i \cos^2 \frac{\theta}{2} & 0 & 0 & i \sin^2 \frac{\theta}{2} \\
        -\frac{m}{\sqrt{s}} \sin\theta & 0 & 0 & \frac{m}{\sqrt{s}} \sin\theta \\
        -\frac{m}{\sqrt{s}} \sin\theta & 0 & 0 & \frac{m}{\sqrt{s}} \sin\theta \\
        -i \sin^2 \frac{\theta}{2} & 0 & 0 & -i \cos^2 \frac{\theta}{2}
    \end{pmatrix},\quad \mathcal{M}_1=\begin{pmatrix}
        i \cos^2 \frac{\theta}{2} & 0 & 0 & i \sin^2 \frac{\theta}{2} \\
        -\frac{\sqrt{s}}{4 m} \sin\theta & 0 & 0 & \frac{\sqrt{s}}{4 m} \sin\theta \\
        -\frac{\sqrt{s}}{4 m} \sin\theta & 0 & 0 & \frac{\sqrt{s}}{4 m} \sin\theta \\
        -i \sin^2 \frac{\theta}{2} & 0 & 0 & -i \cos^2 \frac{\theta}{2}
    \end{pmatrix}.
\end{equation}
Next, we compute the normalization factors using the expressions for $\mathcal{M}_0$, $\mathcal{M}_1$, and the initial density matrix. These normalization factors in Eq.~\eqref{eq:sld:scattering} are found to be: $A_0 = 1/(2c_0)$ and $A_1 = 4$, respectively.

Given the complexity of the full expression for the SLD operator in the MDM scenario, we omit the detailed form here. Instead, after performing the necessary substitutions, we obtain the following collider-accessible, non-entangled SLD operator:
\begin{equation}
    \hat{Q} = 2\left(\frac{s}{4 m^2}-1\right) c_0 \begin{pmatrix}
        -\frac{8m^2}{s} \sin^2\theta & 0 & 0 & 0 \\
        0 & \cos 2\theta + 3 & 0 & 0 \\
        0 & 0 & \cos 2\theta + 3 & 0 \\
        0 & 0 & 0 & -\frac{8m^2}{s} \sin^2\theta
    \end{pmatrix} \,.
\end{equation}
The corresponding QFI is then given by:
\begin{equation}
    F_q^{\rm MDM}(\theta) = \frac{2s \sin^2 \theta \left(\cos2\theta + 3\right) \left(s - 4m^2\right)^2}{m^2 \left(8m^2 \sin^2\theta + s \cos2\theta + 3s\right)^2} \,.
\end{equation}

Finally, we comment the case where the expansion is performed around the Standard Model prediction for the $\tau$ magnetic dipole moment, $a^{\mathrm{SM}}_\tau$, such that
\begin{align}
a_\tau = a^{\mathrm{SM}}_\tau + \Delta a_\tau,
\end{align}
where $\Delta a_\tau$ being the parameter of interest for measurement. Since $a^{\mathrm{SM}}_\tau \ll 1$, we shift the density matrix and scattering amplitude as follows:
\begin{align}
\rho_0 \rightarrow \rho_0 + a^{\mathrm{SM}}_\tau \cdot \rho_1 ~~~ \mathcal{M}_0 \rightarrow \mathcal{M}_0 + a^{\mathrm{SM}}_\tau \cdot \mathcal{M}_1.
\end{align}
The procedure for computing the collider-accessible, non-entangled SLD remains unchanged, and the resulting SLD operator remains diagonal. Therefore, this shift does not affect our main conclusions.

\subsection{EDM}

In the EDM scenario, the unperturbed spin density matrix $\rho_0$ remains identical to that of the MDM case (see Eq.~\eqref{eq:rho0:etau}). The first-order correction $\rho_1$, arising from the small EDM parameter $d =\frac{2m}{e} d_\tau$ (redefined to be dimensionless), is given by:
\begin{equation}
    \rho_1 = -c_0 \sqrt{1 - \frac{4m^2}{s}} \begin{pmatrix}
        0 & \frac{\sqrt{s}}{4 m} \sin 2\theta & -\frac{\sqrt{s}}{4 m} \sin 2\theta & 0 \\
        \frac{\sqrt{s}}{4 m} \sin 2\theta & 0 & -2 i \sin^2\theta & \frac{\sqrt{s}}{4 m} \sin 2\theta \\
        -\frac{\sqrt{s}}{4 m} \sin 2\theta & 2 i \sin^2\theta & 0 & -\frac{\sqrt{s}}{4 m} \sin 2\theta \\
        0 & \frac{\sqrt{s}}{4 m} \sin 2\theta & -\frac{\sqrt{s}}{4 m} \sin 2\theta & 0
    \end{pmatrix}.
\end{equation}
The corresponding amplitude $\mathcal{M}$, expanded to first order in the EDM parameter $d$ is:
\begin{equation}
    \mathcal{M}=\begin{pmatrix}
    i \cos^2 \frac{\theta}{2} & 0 & 0 & i \sin^2 \frac{\theta}{2} \\
    \left(-m - i d \frac{\sqrt{s - 4 m^2}}{4 m} \right) \sin\theta & 0 & 0 & \left(m + i d \frac{\sqrt{s - 4 m^2}}{4 m} \right) \sin\theta \\
    \left(-m + i d \frac{\sqrt{s - 4 m^2}}{4 m} \right) \sin\theta & 0 & 0 & \left(m - i d \frac{\sqrt{s - 4 m^2}}{4 m} \right) \sin\theta \\
    -i \sin^2 \frac{\theta}{2} & 0 & 0 & -i \cos^2 \frac{\theta}{2}
\end{pmatrix}
=\mathcal{M}_0+d \mathcal{M}_1,
\end{equation}
where $M_0$ is the same as in the MDM scenario, and the EDM-specific contribution $M_1$ is given by:
\begin{equation}
    \mathcal{M}_1 = \begin{pmatrix}
        0 & 0 & 0 & 0 \\
        -\frac{i \sqrt{s - 4 m^2}}{4 m} \sin\theta & 0 & 0 & \frac{i \sqrt{s - 4 m^2}}{4 m} \sin\theta \\
        \frac{i \sqrt{s - 4 m^2}}{4 m} \sin\theta & 0 & 0 & -\frac{i \sqrt{s - 4 m^2}}{4 m} \sin\theta \\
        0 & 0 & 0 & 0
    \end{pmatrix} \,.
\end{equation}
The QFI associated with the EDM parameter is computed as:
\begin{equation}
    F_q^{\rm EDM}(\theta) = \frac{2s \sin^2\theta \left(4m^2 - s\right)}{m^2 \left(\cos2\theta \left(4m^2 - s\right) - 4m^2 - 3s\right)}.
\end{equation}

In contrast to the MDM case, no SLD operator can be constructed in the EDM scenario that satisfies the non-entanglement condition. As a result, there is no collider-accessible measurement that can saturate the QFI in the EDM scenario.

Moreover, a direct calculation shows that in Eq.~\eqref{eq:A} $\mathcal{M}_0^\dagger \mathcal{M}_1 = \mathcal{M}_1^\dagger \mathcal{M}_0 = 0$, implying that the SLD operator Eq.~\eqref{eq:sld:scattering} is independent of the initial spin density matrix $\rho_{\rm ini}$. Therefore, adjusting the polarization of the electron and positron beams does not affect the conclusion that no collider-accessible measurements in the EDM scenario can saturate the QFI.

\section{Mathematical Proof of Phase-Space-Restricted CFI Approaching QFI}
\label{sec:CFI-to-QFI}

To obtain non-zero event numbers in collider experiments, one must define quasi-optimal measurements that approximate the true QFI-saturating observables but are localized within finite regions of phase space. We construct such measurements in small neighborhoods around the optimal directions. Let the optimal measurement directions be $\hat{q}^{\mathrm{opt}}_+, \hat{q}^{\mathrm{opt}}_-$, with four corresponding quantum optimal measurement operators:
\begin{equation}
    \{\hat{\Pi}^{\mathrm{opt}}_i\}=\{\hat{E}(\hat{q}^{\mathrm{opt}}_{+},\hat{q}^{\mathrm{opt}}_-),\hat{E}(-\hat{q}^{\mathrm{opt}}_{+},\hat{q}^{\mathrm{opt}}_-),\hat{E}(\hat{q}^{\mathrm{opt}}_{+},-\hat{q}^{\mathrm{opt}}_-),\hat{E}(-\hat{q}^{\mathrm{opt}}_{+},-\hat{q}^{\mathrm{opt}}_-)\} \,.
\end{equation}
Define neighborhoods $\mathcal{C}_1$ and $\mathcal{C}_2$ around $\hat{q}^{\mathrm{opt}}_+$ and $\hat{q}^{\mathrm{opt}}_-$, respectively. 
In the main text, taking MDM case as an example, we choose to expand the neighborhoods by a cone with opening half angle $\delta \theta_\pi$. Thus, the solid angle of phase space region $\mathcal{C}_1$ and $\mathcal{C}_2$ are
\begin{align}
    \text{Area}[\mathcal{C}_{1,2}] = \Delta \Omega_\pm = \Delta \Omega = 2\pi (1 -\cos \delta \theta_\pi).
\end{align}
The four angular combinations above correspond to the  following solid angle phase space 
\begin{align}
 \text{PS} = \{  \mathcal{C}_1 \times \mathcal{C}_2, \quad
    \mathcal{C}_1^- \times \mathcal{C}_2, \quad
    \mathcal{C}_1 \times \mathcal{C}_2^-,  \quad
    \mathcal{C}_1^- \times \mathcal{C}_2^- \}, 
    \label{eq:four-PS}
\end{align}
with  $\mathcal{C}^- \equiv \{ -\hat{q} , | , \hat{q} \in \mathcal{C} \}$. The total covered phase space is then 
\begin{align}
\mathcal{C}_{\text{tot}} = (\mathcal{C}_1 \cup \mathcal{C}_1^-) \times (\mathcal{C}_2 \cup \mathcal{C}_2^-).
\end{align}
The corresponding phase-space–restricted POVM is defined as: 
\begin{equation}
    \{\hat{\Pi}(\hat{q}_+,\hat{q}_-)\}=\{\frac{1}{S}\hat{E}(\hat{q}_+,\hat{q}_-)|\hat{q}_+ \in \mathcal{C}_1\cup\mathcal{C}_1^-, \hat{q}_- \in \mathcal{C}_2\cup\mathcal{C}_2^- \},
\end{equation}
where $S$ is a normalization factor ensuring the POVM completeness:
\begin{align}
\int_{\mathcal{C}_{\text{tot}}} d\Omega_+ d\Omega_- \, \hat{\Pi}(\hat{q}_+, \hat{q}_-) = \mathbb{I}_4.
\end{align}
Using the fact that $\hat{E}$ forms an orthonormal complete basis of measurement operators:
\begin{align}
\hat{E}(\hat{q}_+,\hat{q}_-) + \hat{E}(-\hat{q}_+,\hat{q}_-) + \hat{E}(\hat{q}_+,-\hat{q}_-) + \hat{E}(-\hat{q}_+,-\hat{q}_-) = \mathbb{I}_4,
\end{align}
we obtain:
\begin{align}
    S \cdot \mathbb{I}_4 = \int_{\mathcal{C}_{\text{tot}}} d\Omega_+ d\Omega_-\hat{E}(\hat{q}_+, \hat{q}_-) = \Delta \Omega_+ \Delta \Omega_- \cdot \mathbb{I}_4,
\end{align}
where $\Delta\Omega_{\pm}$ denote the solid angle volumes of $\mathcal{C}_{1,2}$. Thus, we have the normalization factor $S = \Delta \Omega_+ \Delta \Omega_-$.

Given this construction, the normalized classical probability distribution for a spin state $\rho_d$ with respect to the parameter $d$ is:
\begin{equation}
\frac{df_d(\hat{q}+, \hat{q}-)}{d\Omega_+ d\Omega_-} \equiv \mathrm{Tr}[\hat{\Pi}(\hat{q}+, \hat{q}-) \rho_d] = \frac{1}{S}  \mathrm{Tr}[\hat{E}(\hat{q}+, \hat{q}-) \rho_d] \,,
\end{equation}
with the normalization confined to the phase space $\mathcal{C}_{\text{tot}}$.

Expanding the density matrix as $\rho_d = \rho_0 + d \cdot \rho_1 + \mathcal{O}(d^2)$, we also have the expansion of the differential distribution 
\begin{align}
    \frac{df_d(\hat{q}+, \hat{q}-)}{d\Omega_+ d\Omega_-} = \Sigma_0 + d \cdot \Sigma_1 + \mathcal{O}(d^2) \,.
\end{align}
The optimal classical observable is given by:
\begin{equation}
O^{\mathrm{opt}}(\hat{q}+, \hat{q}-) = \frac{\mathrm{Tr}[\rho_1 \hat{E}(\hat{q}+, \hat{q}-)]}{\mathrm{Tr}[\rho_0 \hat{E}(\hat{q}+, \hat{q}-)]}
= \frac{\Sigma_1}{\Sigma_0}.
\end{equation}
The associated CFI becomes:
\begin{align}
F_c = \int_{\mathcal{C}_{\text{tot}}} d\Omega_+ d\Omega_- \, \frac{\Sigma_1^2}{\Sigma_0} =\frac{1}{S} \sum_{i} \int_{\text{PS}_i}d\Omega_{+} d\Omega_{-} \frac{\mathrm{Tr}\left(\rho_1 \hat{E}(\hat{q}_+,\hat{q}_-)\right)^2}{\mathrm{Tr}\left(\rho_0 \hat{E}(\hat{q}_+,\hat{q}_-)\right)}.
\label{eq:CFI}
\end{align}
where $i$ denotes summation over the four angular combinations in solid angle phase space PS from Eq.~\eqref{eq:four-PS}.

By multiplying the SLD condition by the SLD eigenvalue projector $\hat{\Pi}$ and taking the trace, one obtains:
\begin{align}
\mathrm{Tr}(\rho_1 \hat{\Pi}^{\text{opt}}_i) = \lambda_i \, \mathrm{Tr}(\rho_0 \hat{\Pi}^{\text{opt}}_i),    \quad \text{for} \quad i = 1,2,3,4 \,.
\end{align}
Under the vanishing solid angle limit, $\Delta \Omega_{\pm} \to 0$, we have:
\begin{align}
\lim_{\Delta \Omega_{\pm} \to 0} F_c = \sum_i \frac{\mathrm{Tr}(\rho_1 \hat{\Pi}_i^{\text{opt}})^2}{\mathrm{Tr}(\rho_0 \hat{\Pi}_i^{\text{opt}})} = \sum_i \lambda_i^2 \, \mathrm{Tr}(\rho_0 \hat{\Pi}_i^{\text{opt}}) = F_q.    
\end{align}

This completes the proof that the CFI, when evaluated over a restricted solid angle in phase space, asymptotically approaches the QFI as the regions $\mathcal{C}_1$ and $\mathcal{C}_2$ shrink to the optimal measurement directions.
In the case of the CP-violating Higgs scenario, the optimal measurement directions span a transverse plane parameterized by the free azimuthal angles $\phi_{1,2}$, rather than being restricted to four discrete directions. Nevertheless, the same conclusion follows, as a similar derivation applies.

\subsection{Results with Non-Maximal Spin Analyzing Power}
\label{sec:non-maximal-spin-analyzing-power}

In the proof above, we have considered the decay $\tau^\pm \to \pi^\pm \overset{\scriptscriptstyle(-)}{\nu}$, which exhibits maximal spin-analyzing power. In this case, the associated generalized measurement operator is a rank-one projector. More generally, the measurement operator for the decay of a spin-1/2 particle can be expressed as
\begin{align}
    \hat{D}^\pm(\hat{q}_\pm) = \frac{1}{2}\left(\mathbb{I}_2 + \alpha_\pm \times\hat{q}_\pm \cdot \vec{\sigma} \right),
\end{align}
where $\alpha_\pm$ denotes the spin-analyzing power and satisfies $ -1 \leq \alpha_\pm \leq 1$, with $|\alpha_\pm| = 1$ corresponding to maximal spin-analyzing power. 

The joint measurement operator for the biparticle system is given by the tensor product:
\begin{align}
    \hat{E}(\hat{q}_+,\hat{q}_-) = \hat{D}^+(\hat{q}_+) \otimes \hat{D}^-(\hat{q}_-).
\end{align}
Using this operator, the normalized double-differential angular distribution of the pion directions takes the form:
\begin{align}
    \frac{df_d(\hat{q}+, \hat{q}-)}{d\Omega_+ d\Omega_-} 
    = \frac{1}{S}  \mathrm{Tr}[\hat{E}(\hat{q}+, \hat{q}-) \rho_d] 
    = \frac{1}{4S}\sum_{i,j} (1+\alpha_+q_{+}^iB_{i}^{+}+\alpha_-q_{-}^iB_{i}^{-}+\alpha_+\alpha_-q_{+}^iq_{-}^jC_{ij}),
\end{align}
where $B_i^\pm = 0$ in all three scenarios studied, and thus these terms are dropped in the subsequent analysis.

The expansion terms $\Sigma_0$ and $\Sigma_1$ in the differential distribution are:
\begin{align}
    \Sigma_0 = \frac{1}{4S}(1+\alpha_+\alpha_-q_{+}^iq_{-}^jC^{ij}_0), \quad
    \Sigma_1 = \frac{1}{4S}(\alpha_+\alpha_-q_{+}^iq_{-}^j\partial_d C^{ij}_d),
\end{align}
where the spin density matrix is fully characterized by the spin-correlation coefficients $C_{ij}$. The integrand of the CFI, as derived from Eq.~\eqref{eq:CFI}, is:
\begin{align}
    \frac{\Sigma_1^2}{\Sigma_0} = \frac{1}{4S} \times \frac{(\alpha_+\alpha_-q_{+}^iq_{-}^j\partial_d C^{ij}_d)^2}{1+\alpha_+\alpha_-q_{+}^iq_{-}^jC^{ij}_0}.
\end{align}
Neglecting the overall constant, we focus on the second term which depends on the spin analyzing power:
\begin{align}
    \frac{(\alpha_+\alpha_-q_{+}^iq_{-}^j\partial_d C^{ij}_d)^2}{1+\alpha_+\alpha_-q_{+}^iq_{-}^jC^{ij}_0} = |\alpha_+\alpha_-|
    \frac{(q_{+}^iq_{-}^j\partial_d C^{ij}_d)^2}{\frac{1}{|\alpha_+\alpha_-|}+ \frac{\alpha_+\alpha_-}{|\alpha_+\alpha_-|}q_{+}^iq_{-}^jC^{ij}_0}
    \leq 
    \frac{(q_{+}^iq_{-}^j\partial_d C^{ij}_d)^2}{1+ \text{Sign}\left[\alpha_+\alpha_- \right]q_{+}^iq_{-}^jC^{ij}_0} \,,
\end{align}
where we have used $|q_+^i q_-^j C_0^{ij}| \leq 1$. The inequality becomes an equality when $|\alpha_\pm| = 1$.

This result shows that the CFI decreases with decreasing spin-analyzing power, i.e., when $|\alpha_\pm| < 1$. Therefore, achieving the QFI in our three benchmark scenarios requires selecting decay channels with maximal spin-analyzing power, such as $\tau \to \pi \nu$. In this work, we restrict our analysis to such projective measurements and defer the treatment of general POVMs to future studies.

\section{Proof of Equivalence for the Amplitude-Based Method}
\label{sec:proof-of-equivalence}

We begin by assuming the existence of a Hermitian operator $P = P^\dagger$ such that:
\begin{equation}
    \frac{1}{2} P \mathcal{M}_0 - \mathcal{M}_1 + \frac{A_1}{2 A_0} \mathcal{M}_0 = 0 \,.
\end{equation}
Using this assumption, we can rewrite Eq.~(\ref{eq:sld:scattering}) as
\begin{equation}
\frac{1}{2} Q \mathcal{M}_0 - \mathcal{M}_1 + \frac{A_1}{2 A_0} \mathcal{M}_0 = \frac{1}{2} (Q - P) \mathcal{M}_0 \equiv \tilde{Q} \mathcal{M}_0,
\end{equation}
where we define $\tilde{Q} \equiv Q - P$. Since both $Q$ and $P$ are Hermitian, it follows that $\tilde{Q}$ is also Hermitian: $\tilde{Q} = \tilde{Q}^\dagger$. Substituting this into Eq.~\eqref{eq:sld:scattering}, we obtain the condition:
\begin{equation}
\tilde{Q} \mathcal{M}_0 \rho^i \mathcal{M}_0^\dagger + \mathcal{M}_0 \rho^i \mathcal{M}_0^\dagger \tilde{Q} = 0.
\end{equation}

To analyze this further, we multiply both sides of the equation on the left and right by $\tilde{Q}$, yielding:
\begin{align}
    \tilde{Q} \mathcal{M}_0 \rho^i\mathcal{M}_0^\dagger \tilde{Q} = -\mathcal{M}_0 \rho^i \mathcal{M}_0^\dagger \tilde{Q} \tilde{Q} = -\tilde{Q} \tilde{Q} \mathcal{M}_0 \rho^i \mathcal{M}_0^\dagger.
\end{align}
Now, observe that the initial density matrix $\rho^i$ is positive semi-definite. As a result, both $\mathcal{M}_0 \rho^i \mathcal{M}_0^\dagger$ and $\tilde{Q} \mathcal{M}_0 \rho^i \mathcal{M}_0^\dagger \tilde{Q}$ are positive semi-definite as well. Moreover, since $\tilde{Q}^2$ commutes with $\mathcal{M}_0 \rho^i \mathcal{M}_0^\dagger$, the operator $\tilde{Q}^2 \mathcal{M}_0 \rho^i \mathcal{M}_0^\dagger$ is also positive semi-definite.

However, the equality above shows that a positive semi-definite matrix equals its negative. This can only be satisfied if the matrix itself is zero. Therefore,
\begin{equation}
    \tilde{Q} \mathcal{M}_0 \rho^i \mathcal{M}_0^\dagger \tilde{Q} = 0.
\end{equation}
Assuming further that the initial state $\rho^i$ is full-rank, a common scenario in unpolarized systems such as $e^+e^-$ collisions, this leads directly to:
\begin{equation}
    \tilde{Q} \mathcal{M}_0 = 0.
\end{equation}
This demonstrates that Eq.~\eqref{eq:sld:scattering} is equivalent to Eq.~\eqref{eq:sufficient} under the stated assumptions.

Finally, we comment on the existence of such a Hermitian operator $P$. If the necessary conditions (1) and (2) in Eq.~\eqref{eq:sufficient} are satisfied, one can explicitly construct a Hermitian solution for $Q$ as:
\begin{equation}
    Q = \tilde{\mathcal{M}}_1 \mathcal{M}_0^{\mathrm{MP}} + (\tilde{M}_1 \mathcal{M}_0^{\mathrm{MP}})^\dagger - \mathcal{M}_0 \mathcal{M}_0^{\mathrm{MP}} \tilde{\mathcal{M}}_1 \mathcal{M}_0^{\mathrm{MP}} + (I - \mathcal{M}_0 \mathcal{M}_0^{\mathrm{MP}}) V (I - M_0 \mathcal{M}_0^{\mathrm{MP}}),
\end{equation}
where $\tilde{\mathcal{M}}_1 = \mathcal{M}_1 - \frac{A_1}{2 A_0} \mathcal{M}_0$, $\mathcal{M}_0^{\mathrm{MP}}$ is the Moore-Penrose pseudoinverse of $\mathcal{M}_0$, and $V$ is an arbitrary Hermitian operator. This construction ensures the Hermiticity of $Q$, thereby validating the equivalence.

\nocite{Ye:2024dqx}

\bibliography{mainrefs}
\bibliographystyle{JHEP}

\end{document}